\documentclass[12pt, preprint]{emulateapj}
\usepackage[]{natbib}
\shorttitle{GC color magnitude relations}
\shortauthors{Mieske et al.}
\def\etal{{\it et~al.}~}

\def\kmm{{\tt KMM~}}

\begin{document}

\title{The ACS Virgo Cluster Survey. XIV. Analysis of Color-Magnitude
Relations in Globular Cluster Systems\altaffilmark{1}}

\author{Steffen Mieske\altaffilmark{2}, Andr\'es
Jord\'an\altaffilmark{2,3}, Patrick C\^{o}t\'{e}\altaffilmark{4},
Markus Kissler-Patig\altaffilmark{2}, Eric W. Peng\altaffilmark{4},
Laura Ferrarese\altaffilmark{4}, John P. Blakeslee\altaffilmark{5,6},
Simona Mei\altaffilmark{5}, David Merritt\altaffilmark{7}, John
L. Tonry\altaffilmark{8}, \& Michael J. West\altaffilmark{9}}
\altaffiltext{1}{Based on observations with the NASA/ESA {\it Hubble
Space Telescope} obtained at the Space Telescope Science Institute,
which is operated by the association of Universities for Research in
Astronomy, Inc., under NASA contract NAS 5-26555.}
\altaffiltext{2}{European Southern Observatory,
Karl-Schwarzschild-Strasse 2, 85748 Garching bei M\"unchen, Germany;
{\sf smieske@eso.org, ajordan@eso.org, mkissler@eso.org}}
\altaffiltext{3}{Astrophysics, Denys Wilkinson Building, University of
Oxford, 1 Keble Road, OX1 3RH, UK} \altaffiltext{4}{Herzberg Institute
of Astrophysics, Victoria,  BC V9E 2E7, Canada; {\sf
patrick.cote@nrc-cnrc.gc.ca, laura.ferrarese@nrc-cnrc.gc.ca,
eric.peng@nrc-cnrc.gc.ca}} \altaffiltext{5}{Department of Physics and
Astronomy, The Johns Hopkins University, 3400 North Charles Street,
Baltimore, MD 21218-2686; {\sf jpb@pha.jhu.edu, smei@pha.jhu.edu}}
\altaffiltext{6}{Department of Physics and Astronomy, PO Box 642814,
Washington State University, Pullman, WA 99164}
\altaffiltext{7}{Department of Physics, Rochester Institute of
Technology, 84 Lomb Memorial Drive, Rochester, NY 14623; {\sf
merritt@cis.rit.edu}} \altaffiltext{8}{Institute for Astronomy,
University of Hawaii, 2680 Woodlawn Drive, Honolulu, HI 96822; {\sf
jt@ifa.hawaii.edu}} \altaffiltext{9}{Gemini Observatory
Southern Operations Center
c/o AURA, Casilla 603
La Serena, Chile; {\sf
mwest@gemini.edu}}

\begin{abstract}
\noindent We examine the correlation between globular cluster (GC)
color and magnitude using HST/ACS imaging for a sample of 79 early-type  
galaxies ($-21.7<M_B<-15.2$ mag) with accurate surface-brightness
fluctuation distances from the ACS Virgo Cluster Survey.
Using the \kmm mixture modeling algorithm, we find a highly
significant correlation,
$\gamma_z \equiv \frac{d(g-z)}{dz} = -0.037 \pm 0.004$, between
color and magnitude for the subpopulation of blue GCs in
the co-added GC color-magnitude diagram of the
three brightest Virgo cluster galaxies (M49, M87 and M60).
The sense of the correlation is such that brighter GCs are redder than
their fainter counterparts. For the single GC systems of M87 and M60, we find similar
correlations; M49 does not appear to show a significant trend. There
is no correlation between $(g-z)$ and $M_z$ for GCs belonging to the red
subpopulation.  The correlation $\gamma_g\equiv \frac{d(g-z)}{dg}$ for the blue
subpopulation is much weaker than $\gamma_z$. Using
Monte Carlo simulations, we attribute this finding to the
fact that the blue subpopulation in $M_g$ extends to higher
luminosities than does the red subpopulation, which biases the
\kmm fit results. The highly significant correlation between
color and $M_z$, however, is a real effect: this conclusion is 
supported by biweight fits to the same color distributions. We identify
two environmental dependencies which influence the derived color-magnitude
relation: (1) the slope of the color-magnitude relation decreases in
significance with decreasing galaxy luminosity, although it remains
detectable over the full luminosity range of our sample; and (2)
the slope is stronger for GC populations located at smaller
galactocentric distances. These characteristics suggest that the
observed trend is, at least partially, shaped by external agents. We
examine several physical mechanisms that might give rise to the
observed color-magnitude relation including: (1) presence of
contaminants like super-clusters, stripped galactic nuclei, or 
ultra-compact dwarfs; (2) accretion of GCs from low-mass 
galaxies;  (3) stochastic effects; 
(4) the capture of
field stars by individual GCs; and (5) GC self-enrichment. Although none of these scenarios
offers a fully satisfactory explanation of the observations, we 
conclude that self-enrichment and field-star capture, or a combination
of these processes, offer the most promising means of explaining
our observations.
\end{abstract}


\section{Introduction}
It is by now an accepted paradigm that the formation of globular
clusters (GCs) is closely linked to the formation of their host
galaxies (e.g. Searle \& Zinn~1978, Ashman \&
Zepf~1992, Kissler-Patig~1997, Hilker {\it et al.}~1999b, 
Forbes \etal~1997, C\^{o}t\'{e} {\it et
al.}~1998, Beasley \etal~2002). In recent years,
observational data for extragalactic GC systems
have made a huge leap forward in quality,
thanks mainly to high-resolution imaging from the 
Hubble Space telescope, and to wide field imaging and multi-object spectroscopy
from 8m-class ground based telescopes (see West \etal~2004 and references therein).

As a rule, studies of extragalactic GCs have tended to focus on 
correlations between the global photometric properties of the GC systems
(GCSs) and those of their host galaxy. For instance, Peng \etal~\cite{Peng05},
examined the color distributions for GCs belonging to 100 galaxies 
observed in the
ACS Virgo Cluster Survey (ACSVCS; C\^ot\'e \etal~2004; see also below), 
finding 
nearly all of these galaxies ($-22<M_B<-15$ mag) to possess bimodal,
or at least asymmetric, GC color distributions. Consistent with
previous findings (e.g., Gebhardt \& Kissler-Patig~1999; Kundu \etal~2001;
Larsen \etal~2001), the mean GC color was found to correlate with galaxy 
luminosity in the same sense as the well-known color-magnitude
relation for early type galaxies (e.g. Bower, Lucey \&
Ellis~1992; Karick \etal~2003; Hilker \etal~2003; Ferrarese \etal~2006).
A corresponding GC color - galaxy color relation was also found by Peng
\etal~\cite{Peng05}, confirming previous findings 
(e.g., Burgarella \etal~2001; Larsen \etal~2001; Lotz 
\etal~2004).

At the same time, analyses of the color magnitude diagrams for the GCSs
of early-type galaxies revealed some tentative evidence for a
dependence of {\it individual} GC color on
luminosity.  Ostrov \etal~\cite{Ostrov98} found from
Washington photometry of NGC 1399 that the color peaks of the GC color
distribution seem to merge at the highest luminosities. Dirsch et
al.~\cite{Dirsch03} confirmed this finding for the same galaxy based
on an expanded sample of GCs. There have also been hints of such
a behavior in other galaxies such as NGC 5128 (Harris \etal~2004) and M87 
(Whitmore \etal~1995; Larsen \etal~2001).

With the improved imaging capabilities of HST provided by the Advanced
Camera for Surveys (ACS; Ford \etal~1998), it has become possible to
investigate this issue in more detail.  Harris \etal~\cite{Harris06}
used ACS imaging to investigate the distribution of GCs in the $(I,
B-I)$ color magnitude plane for eight brightest cluster galaxies in
the range $1800 \lesssim cz \lesssim 3200$ km s$^{-1}$. For the four
galaxies with the largest number of GCs --- and also the joint sample
of eight GC systems --- they found a significant trend towards redder
colors for increasing luminosity for the subpopulation of blue GCs.
This finding led them to suggest a mass-metallicity relation of the
form $Z \propto M^{0.55}$ for this subpopulation.  Curiously, no
color-magnitude relation was found for the red GCs.

Strader \etal~\cite{Strade05} used the publicly available data from the
ACSVCS to look for color-magnitude trends in the three brightest Virgo
Cluster galaxies M49, M87 and M60. For the blue GC populations of M87
and M60, they found a significant trend of redder color with brighter
GC $z$-band luminosity similar to that found by Harris \etal~\cite{Harris06}.
They too interpreted this trend as evidence for a mass-metallicity
relation among the blue (metal-poor) GCs and suggested self-enrichment
as a possible explanation. No significant color-magnitude trend
was found for M49, the brightest member of the Virgo Cluster.

Both the Harris \etal~\cite{Harris06} and Strader
\etal~\cite{Strade05} studies used heteroscedastic mixture models to
fit the colors of an assumed double-Gaussian color distribution to the
observed GC color-magnitude diagrams (CMDs) as a function of
magnitude. Harris {\it et~al.} use {\kmm} (e.g. Ashman, Bird \&
Zepf~1994), while Strader {\it et~al.} use {\tt Nmix} (Richardson \&
Green~1997).  Two important issues that were not addressed by these
studies are the subject of this paper: first, we will investigate the
color-magnitude correlations not only in the red (i.e., $z$), but also
in the blue band (i.e., $g$); second, we will investigate how the blue
peak slope varies as a function of host galaxy luminosity and
galactocentric distance. Note that the Harris \etal~\cite{Harris06}
study targeted only brightest cluster galaxies, so it spanned a
relatively limited range in galaxy luminosity.

We finally point out that there is recent evidence for the existence
of a GC color-magnitude relation also in spiral galaxies. Based on ACS
imaging, Spitler {\it et~al.}~\cite{Spitle06} found a significant
color-magnitude trend for the Sombrero galaxy (NGC 4594) implying $Z
\propto M^{0.30}$, which is a somewhat weaker trend than suggested by
Harris \etal~\cite{Harris06} and Strader \etal~\cite{Strade05} for
giant ellipticals.

The aim of this paper is to investigate the color-magnitude trend of
GCs over the entire host galaxy luminosity range covered by the
ACSVCS. Special emphasis will be put on environmental dependencies of
this trend and checks of the fitting technique. The paper is
structured as follows: In \S\ref{gcsel} we outline how GC
candidates are selected in the ACSVCS. In \S\ref{realvssim} we
compare \kmm fits to simulated CMDs with fits to observed CMDs. In
\S\ref{environment} we investigate the environmental dependence of
the color-magnitude trend. In \S\ref{discussion} we examine a
variety of physical effects that may create these trends. We finish
this paper with the summary and conclusions in \S\ref{conclusions}.

\section{Selection of Globular Clusters from the ACSVCS}
\label{gcsel}

The ACSVCS represents the most complete and homogeneous study of
extragalactic GCs performed to date. The sample consists of 100
early-type (E, S0, dE, dE,N, dS0) members of the Virgo cluster.  Each
galaxy was imaged in the F475W and F850LP filters ($\approx g_{475}$
and $z_{850}$, respectively) for a total of 750 seconds and 1210
seconds, respectively. This filter combination gives roughly a
factor-of-two improvement in wavelength baseline and metallicity
sensitivity compared to the ``canonical'' $(V-I)$ color index. The
identification of bona-fide GCs from these images is performed in the
size-magnitude plane as described in Peng \etal\cite{Peng05}. This
selection procedure --- which is possible because the half-light radii
of GCs are marginally resolved (Jord\'an \etal\cite{Jordan05a}) at the
distance of Virgo ($d =16.5$ Mpc; Tonry \etal\cite{Tonry01}; Mei
\etal\cite{Mei05}) --- greatly reduces contamination from foreground
stars and background galaxies.

The ACS images had been reduced using a dedicated pipeline described
in detail by Jord\'an \etal (2004ab; Papers~II and III).  In brief,
the reduction steps are: image combination, galaxy modeling and model
subtraction, rejection of obvious background galaxies, and the
measurement of magnitudes and sizes for candidate GCs using the
program KINGPHOT (Jord\'an \etal 2005; Paper X).  The result is a
catalog of integrated $g$ and $z$ magnitudes, $(g-z)$ colors and
half-light radii, $r_h$, for each candidate GC. Magnitudes and colors
are corrected for foreground extinction using the reddening maps of
Schlegel \etal~\cite{Schleg98}.

To estimate the contamination by background galaxies, an identical
reduction procedure was applied to 17 blank high-latitude fields
observed by ACS in the $g$ and $z$ filters. Because background
galaxies are typically larger at comparable luminosities (or,
conversely, fainter at comparable sizes), a statistical
decontamination is performed in size-magnitude space (see Peng
\etal~2006 and Jord\'an \etal 2006, in prep. for details): to this
end, first a non-parametric density model is fitted to the
size-magnitude distribution of sources detected in background fields.
For the GC candidate magnitude distribution, a Gaussian luminosity
function is assumed. The GC candidate size distribution is fitted by a
non-parametric kernel estimate plus a power law tail.  The resulting
surface densities in the size-magnitude plane for background sources
and GCs then make it possible to assign a GC probability, $\cal P_{\rm
  gc}$, to every source detected in the galaxy image. All candidates
with probabilities ${\cal P}_{\rm gc} \ge 0.5$ are considered GCs.
The remaining contamination by background galaxies that are assigned
${\cal P}_{\rm gc} \ge 0.5$ is negligible for the sake of this study.
We have tested this by statistically cleaning the GC CMDs for each
galaxy, using the color-magnitude distribution of sources with ${\cal
  P}_{\rm gc} \ge 0.5$ in the 17 blank background fields.
 
\begin{figure}
\plotone{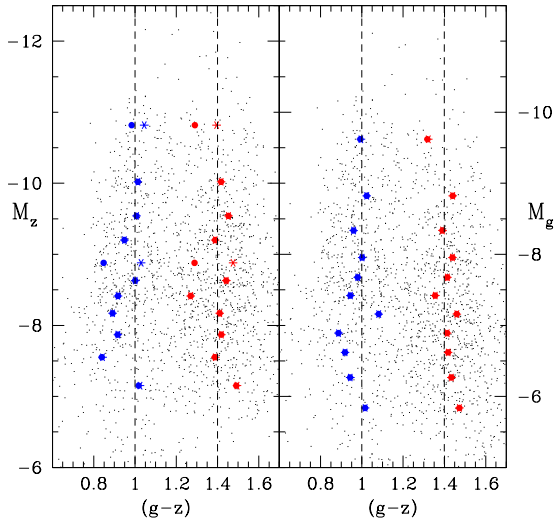}
\caption{\label{CMD1316} Color-magnitude diagram for GCs in M87 (= NGC4486 = VCC1316) 
  using $M_z$ {\it (left panel)} and $M_g$ {\it (right panel)}. Dashed
  lines at $(g-z)=1.0$ and $(g-z)=1.4$ are shown as a visual aid; they
  correspond to the approximate median colors of the red and blue
  subpopulations.  Circles and asterisks indicate \kmm fitting
  results. Circles show \kmm results for initial guesses of 0.8 and
  1.2~mag for the two peaks; asterisks show the results obtained for
  initial guesses of 1.0 and 1.4~mag.}
\end{figure}

\begin{figure*}
\centerline{\plottwo{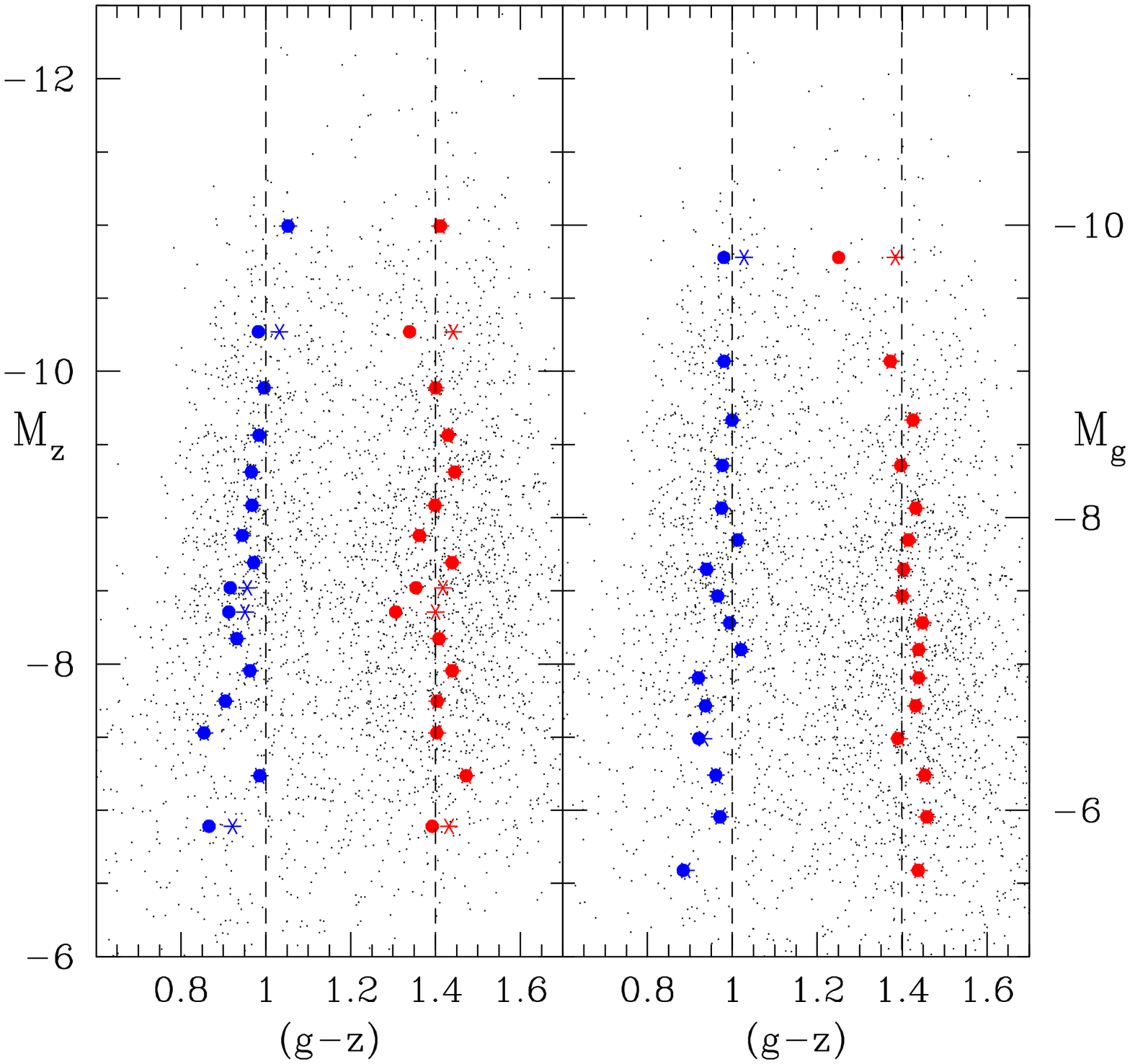}{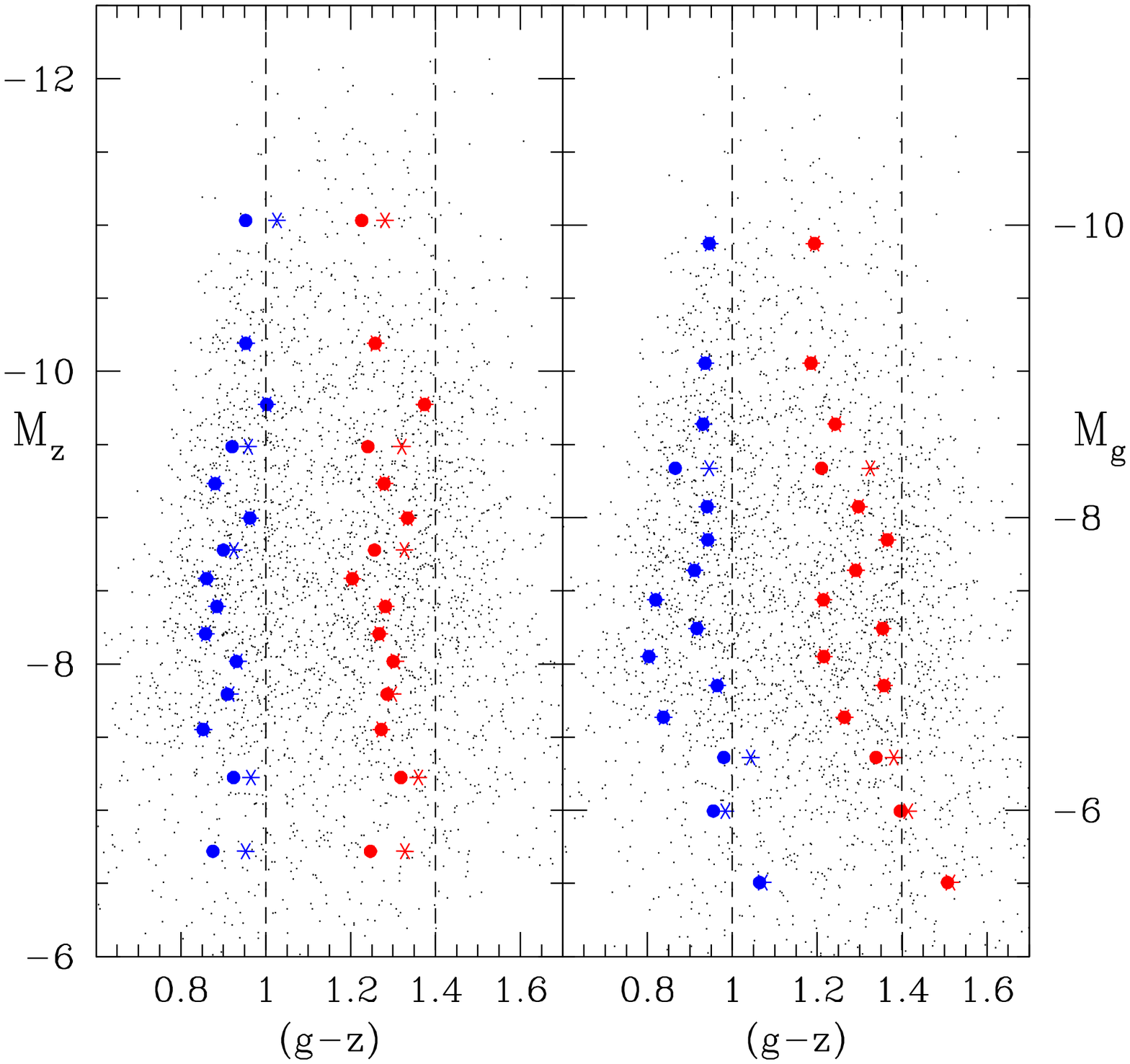}}
\centerline{\plottwo{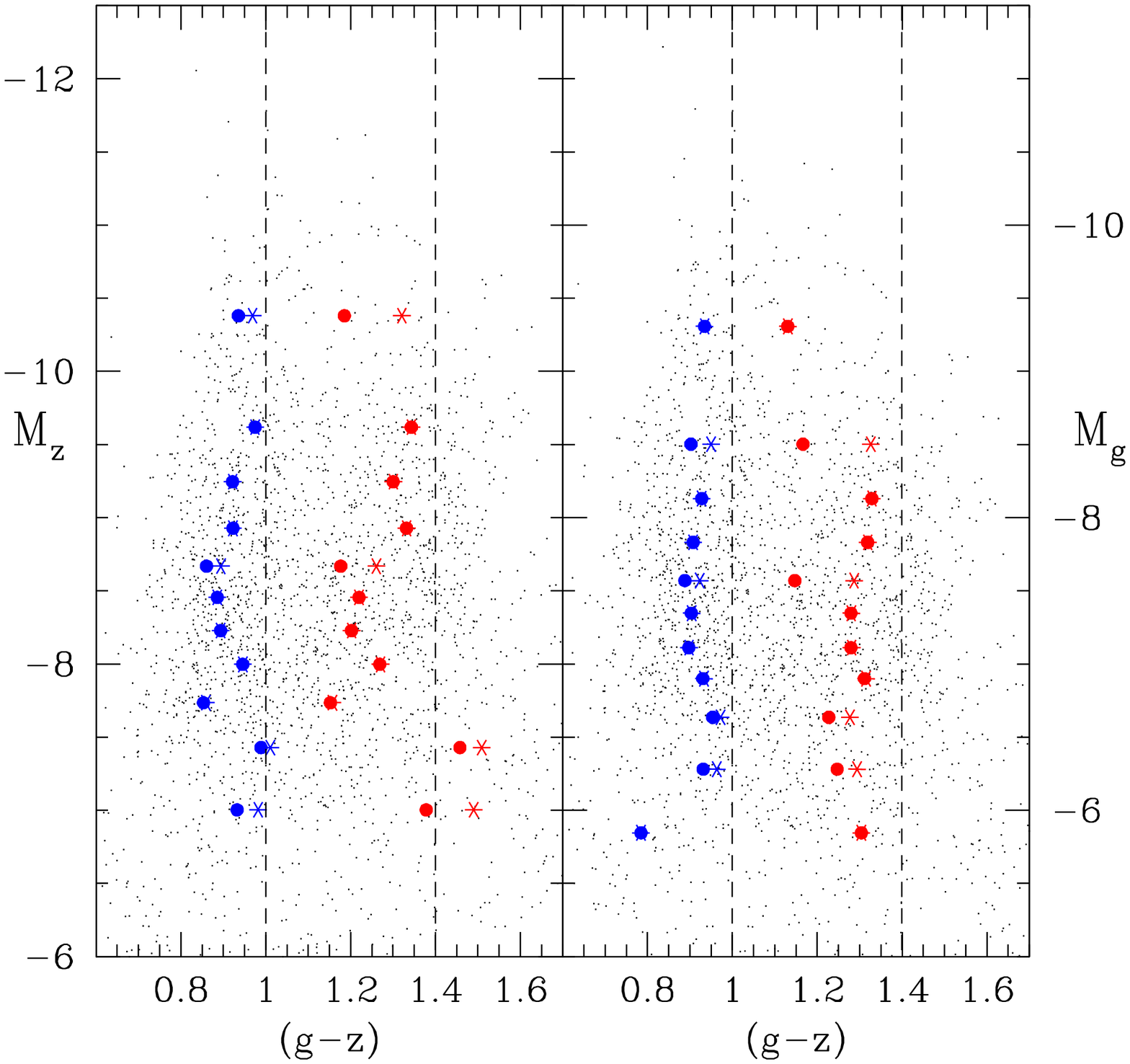}{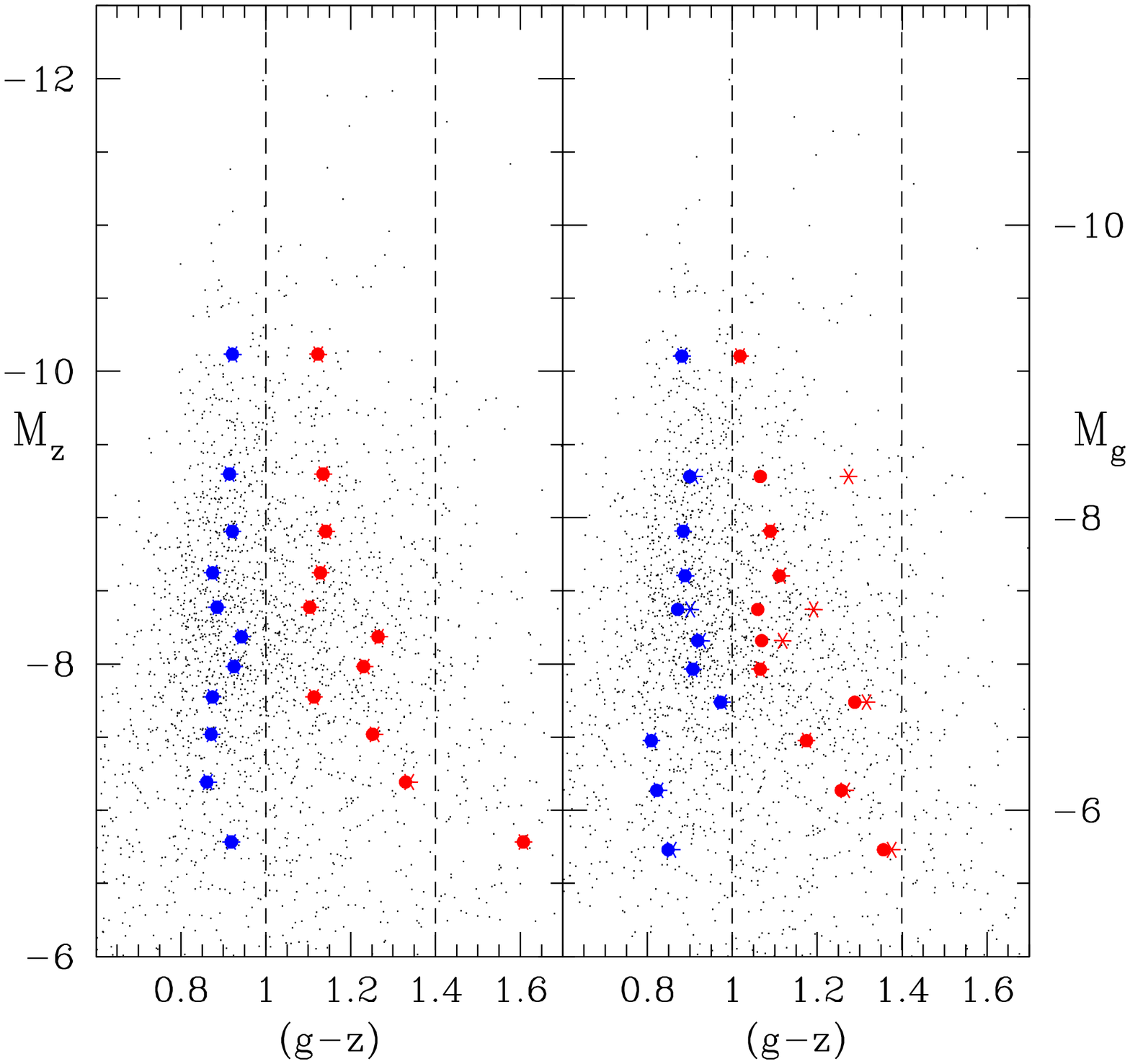}}
\caption{\label{CMDclasses}The same as Figure~\ref{CMD1316}, except for the combined
  sample of GCs in each of the four host-galaxy luminosity groups (see
  Table~\ref{Fitresults}).  {\it (Upper left)} The co-added CMD for
  GCs belonging to the three brightest ACSVCS galaxies, having
  $M_B<-21$ mag (Group 1).  {\it (Upper right)} The co-added CMD for
  GCs belonging to the six ACSVCS galaxies with $-21<M_B<-20$ mag
  (Group 2).  {\it (Lower left)} The co-added CMD for GCs belonging to
  the 16 ACSVCS galaxies with $-20<M_B<-18.4$ mag (Group 3).  {\it
    (Lower right)} The co-added CMD for GCs belonging to 53 ACSVCS
  galaxies with $-18.4<M_B<-15.2$ mag (Group 4). The luminosity groups
  have been chosen such that the number of GCs in each co-added CMD
  are comparable (i.e., $N_{\rm gc} = 2400-3300$).}
\end{figure*}

\section{Analysis of the Color Magnitude Diagrams}
\label{realvssim}
\subsection{Results from KMM}
Figure~\ref{CMD1316} shows $M_g$-$(g-z)$ and $M_z$-$(g-z)$ CMDs for
the GCS of M87 (= NGC4486 = VCC1316), the central galaxy of the Virgo
Cluster and the galaxy with the singlest largest population of GCs.
For comparison, Figure~\ref{CMDclasses} shows the co-added CMDs for
the 79 ACSVCS galaxies that have reliable SBF distances (Mei \etal
2005, 2006) and which have a discernible GC population (see Peng \etal
2006).  We use the SBF distance moduli from Mei \etal~(2005, 2006),
which have an internal precision of $\approx 0.07$~mag, to assign
absolute magnitudes to the associated GC candidates. To better
visualize trends across host galaxy magnitude, the 79 galaxies are
subdivided into four broad luminosity groups such that the number of
GCs per group falls in the range $N_{\rm gc} \approx$ 2400 to 3300.

Luminosity Group 1 consists of the three brightest AVCVCS galaxies M49
(= NGC4472 = VCC1226; $M_B=-21.7$ mag), M87 (= NGC4486 = VCC1316;
$M_B=-21.4$ mag), and M60 (= NGC4649 = VCC1978; $M_B=-21.2$ mag).
Group 2 consists of six ACSVCS galaxies with $-21<M_B<-20$~mag.  Group
3 consists of 17 ACSVCS galaxies with $-20<M_B<-18.4$~mag.  The
remaining 53 ACSVCS galaxies with $-18.4<M_B<-15.2$~mag make up Group
4.  The three galaxies in Group 1 are all so-called ``core-S\'ersic
galaxies'' (Ferrarese \etal 2006), while those in Groups 2 and 3 are a
mixture of core-S\'ersic and S\'ersic elliptical and lenticular
galaxies.  The 53 galaxies in Group 4 consist mainly, but not
exclusively, of early-type dwarfs (e.g., dE, dE,N, dS0 and dS0,N).

We have applied \kmm to the combined CMDs for these four groups, as
well as to the individual CMDs for the three galaxies in Group 1. To
this end, we subdivided the CMDs into luminosity bins, each bin
containing the same number of data points. For the CMDs of M49 and
M60, the size of each luminosity bin is 100, for M87 the size is 150,
and for the combined CMDs the bin size is 200 GCs. \kmm was then run
on the GC color distribution in each luminosity bin. As in the paper
by Harris {\it et~al.}~\cite{Harris06}, we use the heteroscedastic option of
{\tt KMM}: we allow for Gaussians of different widths to be fitted to
the blue and red GC populations. The fitted mean positions are plotted
over the respective CMDs in Figures~\ref{CMD1316} and
\ref{CMDclasses}.  We show the \kmm results for two different initial
guesses for the mean colors of the blue and red GC populations: one
for the pair $(g-z)=0.8$ and 1.2~mag, another one for $(g-z)=1.0$ and
1.4~mag.  These different guesses were chosen to cover: (1) the range
of peak positions between CMDs of the various luminosity groups; and
(2) the range caused by the color-magnitude trend itself (see
Figure~\ref{CMDclasses}).  Linear slopes between color and absolute
magnitude are obtained from ordinary least square fits to the \kmm
peak positions for both pairs of initial guesses, fitting color as a
function of magnitude. GCs fainter than $M_z=-7.7$~mag and
$M_g=-6.7$~mag were excluded from the fits, to avoid effects of
photometric incompleteness on the measured slopes.  The mean of both
fits is adopted as the final best-fit slope. The fit errors of the
slope are derived from re-sampling the points using as dispersion the
observed scatter around the fitted relation. We note that this scatter
is for all luminosity groups larger (by 30\% for luminosity group 1
and by 100\% for group 4) than the average statistical error estimate
for the peak position, given by $\frac{\sigma}{\sqrt{N}}$.  Here,
$\sigma$ is the \kmm width of the Gaussian peak and $N$ the average
number of GCs per magnitude bin associated to the peak. We are
therefore confident that our error estimates are not too optimistic.
For the few cases where the difference between the results obtained
using the two initial guesses was larger than the formal fit errors,
we adopted the difference as the error. The resulting slopes of those
fits are shown in Table~\ref{Fitresults}.

We first consider $M_z$ vs. $(g-z)$ for Group 1. In this case, we
detect a very significant slope $\gamma_z \equiv
\frac{d(g-z)}{dz}=-0.037 \pm 0.004$ for the blue peak position. When
restricting the fit to $M_z>-10$ mag (i.e., rejecting the two
brightest magnitude bins in Figure~\ref{CMDclasses}), the result is
$\gamma_z = -0.033 \pm 0.007$. This is significant at the
4.5$\sigma$ level, indicating that the color-magnitude trend is {\it
  not} restricted to just the brightest GCs.  For M87 and M60 alone, a
comparable slope is derived, albeit at somewhat lower ($\approx
3\sigma$) significance. The slope for M49 is consistent with zero, in
agreement with the findings of Strader \etal~\cite{Strade05}.  When
excluding M49 from Group 1, the slope for the blue population rises
slightly to $\gamma_z \approx -0.040$. The non-detection of a
color-magnitude trend for M49 is intriguing, since this galaxy has a
similar number of GCs as M60 and is similar to M60 and M87 in terms of
luminosity, color, surface brightness and GC dynamics (Peng \etal
2006; Ferrarese \etal 2006; Mei \etal~2005; C\^ot\^e \etal 2001;
2003). The fact that M49 is located at the center of its own
sub-cluster --- being offset by $\approx 4.5^{\circ}$ from the Virgo
Cluster center --- may hint at some environmental effect on the
strength of the observed color-magnitude trend.

In contrast to the strong correlation $\gamma_z$ exhibited by the blue
GCs in Group 1, $\gamma_z$ is insignificant for the red GCs in this
group.  However, the situation changes when using $M_g$ instead of
$M_z$ as magnitude in the color-magnitude diagram (see
Table~\ref{Fitresults}). The slope $\gamma_g \equiv \frac{d(g-z)}{dg}$
for the blue population is more than 50\% smaller than $\gamma_z$ and,
as a result, has a lower level of significance (1.7$\sigma$.)
Considered individually, $\gamma_g$ of the blue population becomes
insignificant for all three galaxies in this group. In turn, the slope
$\gamma_g$ of the red population becomes marginally positive --- i.e.,
opposite in sense to that defined by $\gamma_z$ for the blue
population.

\begin{figure*}
\plottwo{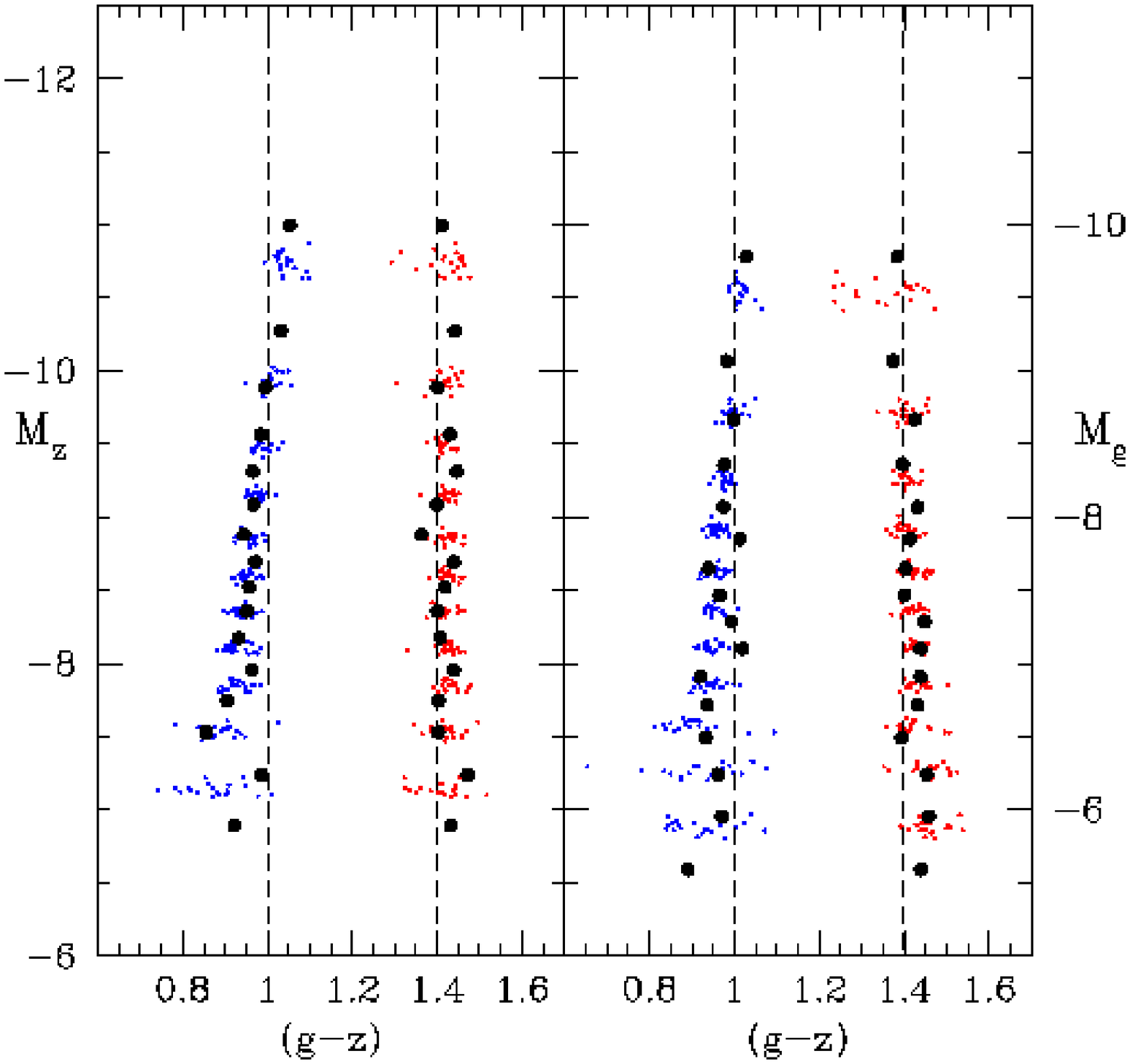}{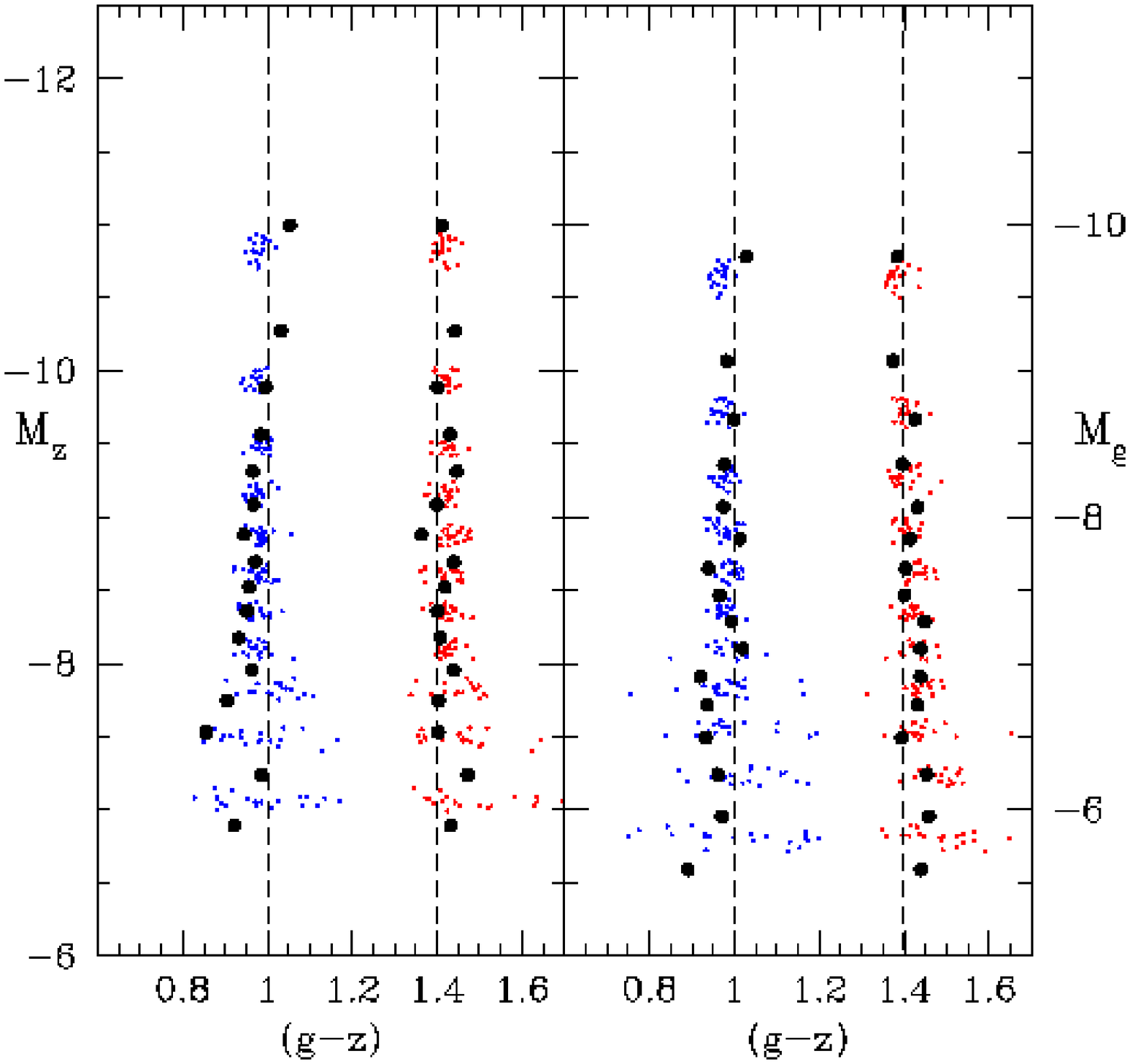}
\caption{\label{CMD3brightsim} \kmm result for the co-added CMD of the three 
  brightest ACSVCS galaxies (Group 1, see Figure~\ref{CMDclasses} and
  Table~\ref{Fitresults}) compared to \kmm results for simulated M87
  CMDs. Filled circles indicate \kmm fitting results for the real data
  with initial guesses of 1.0 and 1.4 mag for peak positions (see
  Figures~\ref{CMD1316} and~\ref{CMDclasses}). Dots indicate \kmm peak
  fittings for 20 simulated CMDs.  {\it (Left Panel)} Simulated CMDs
  with a slope of $\gamma_z \equiv \frac{d(g-z)}{dz} =-0.04$
  implemented for the blue subpopulation; no correlation has been
  implemented for the red subpopulation.  Note the artificial bending
  of the red sequence toward blue colors at bright luminosities,
  especially for $M_g$. {\it (Right Panel)} Simulated CMD with {\it
    no} color-magnitude relation implemented for either population.}
\end{figure*}

\begin{figure*}
\plottwo{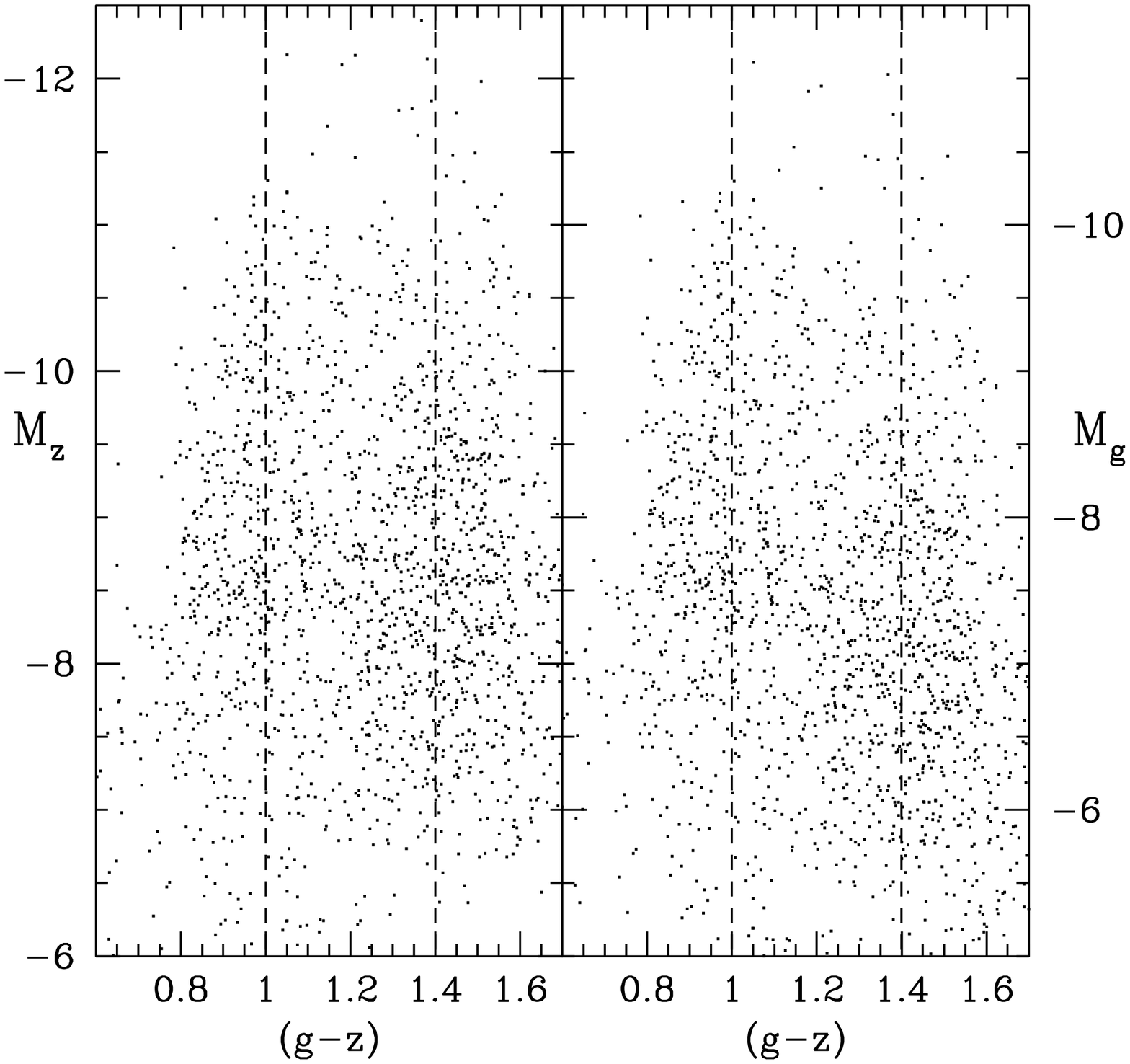}{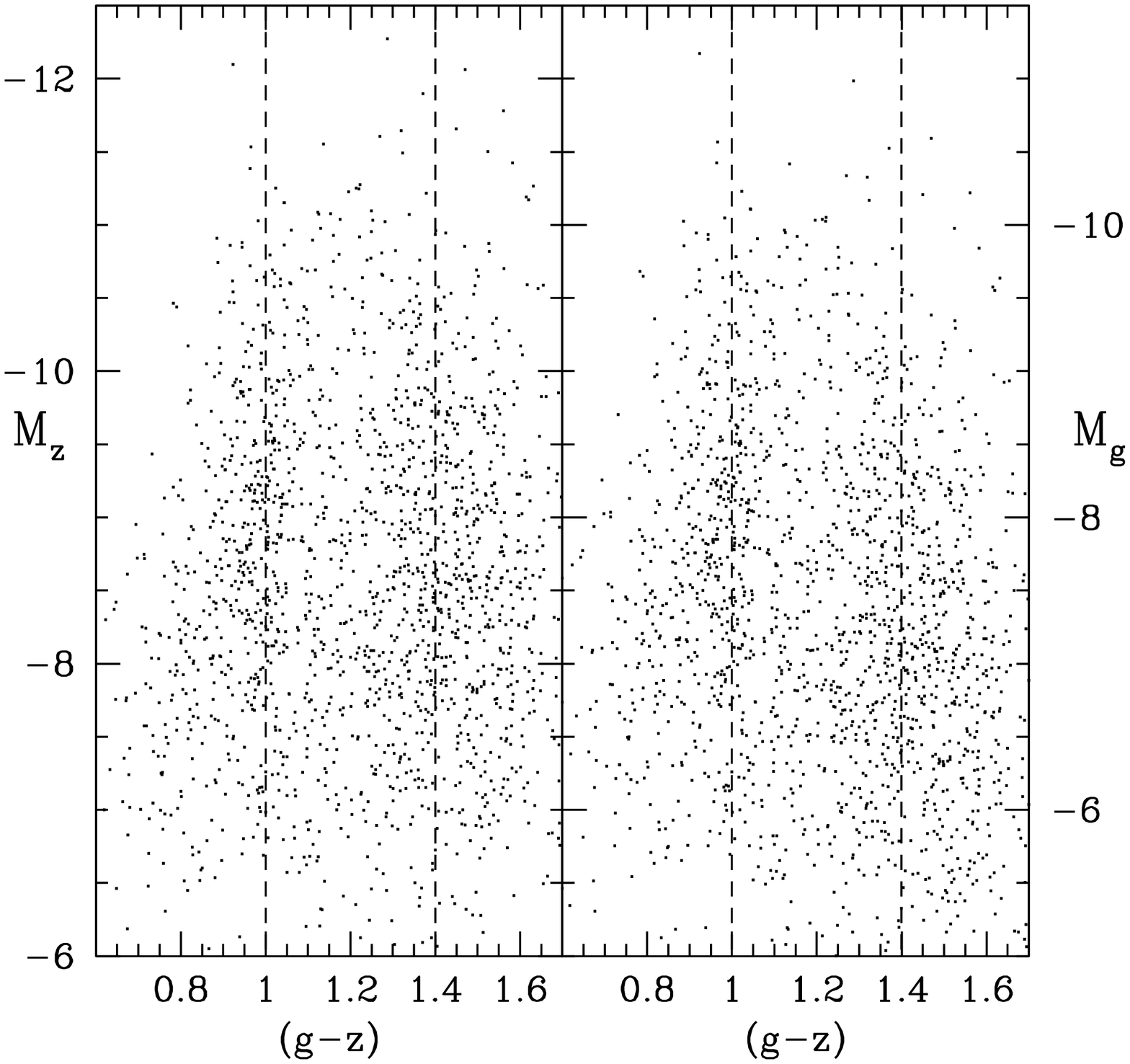}
\caption{\label{Compsim}Comparison between real ({\it left panel}) and simulated 
  ({\it right panel}) CMD for GCs in M87. The simulated CMD has a
  slope of $\gamma_z=-0.04$ input for the blue subpopulation.}
\end{figure*}

\begin{table*}
\caption{Color-magnitude trends for red and blue globular clusters in 
ACS Virgo Cluster Survey galaxies, determined with \kmm fits\label{Fitresults}}
\small
\begin{center}
\begin{tabular}{|l|rrrr|}
\hline\hline Sample &  $\gamma_{z,\rm blue}$ &  $\gamma_{z,\rm red}$ &  
$\gamma_{g,\rm blue}$ & $\gamma_{g,\rm red}$\\\hline M49 ($M_B=-21.7$)&   -0.008 $\pm$ 0.024&0.013 $\pm$
0.018& 0.021 $\pm$ 0.018&0.043 $\pm$ 0.008\\ M87 ($M_B=-21.4$)& -0.042
$\pm$ 0.015&0.003 $\pm$ 0.023& -0.016 $\pm$ 0.023&0.025 $\pm$ 0.017\\
M60 ($M_B=-21.2$)&   -0.028 $\pm$ 0.009&0.012 $\pm$ 0.018& -0.003
$\pm$ 0.033&0.072 $\pm$ 0.017\\\hline Simulated M87 NoCM&0.002 $\pm$
0.012&0.010 $\pm$ 0.015&0.008 $\pm$ 0.015&0.028 $\pm$ 0.022\\
Simulated M87 CM&-0.039 $\pm$ 0.009&0.010 $\pm$ 0.014&-0.028 $\pm$
0.010&0.031 $\pm$ 0.024\\\hline Group1 ($-21.7<M_B<-21$)& -0.037 $\pm$
0.004& -0.001 $\pm$ 0.007& -0.015 $\pm$ 0.009& 0.032 $\pm$ 0.019\\
Group2 ($-21<M_B<-20$)& -0.033 $\pm$ 0.011& 0.001 $\pm$ 0.014& -0.022
$\pm$ 0.016& 0.042 $\pm$ 0.018\\ Group3 ($-20<M_B<-18.4$)& -0.032
$\pm$ 0.012& -0.043 $\pm$ 0.024& -0.010 $\pm$ 0.007& 0.055 $\pm$
0.029\\ Group4 ($-18.4<M_B<-15.2$)& -0.009 $\pm$ 0.013& 0.028 $\pm$
0.028& 0.025 $\pm$ 0.012& 0.058 $\pm$ 0.040\\ Group1, inner
regions&-0.061 $\pm$ 0.014&-0.006 $\pm$ 0.014& -0.007 $\pm$ 0.018 &
0.059 $\pm$ 0.018\\ Group1, outer regions&-0.026 $\pm$ 0.008&0.003
$\pm$ 0.020& -0.004 $\pm$ 0.015&0.050 $\pm$ 0.014\\ Group1+2, inner
regions&-0.066 $\pm$ 0.007& -0.024 $\pm$ 0.006& -0.022 $\pm$
0.020&0.043 $\pm$ 0.018\\ Group1+2, outer regions&-0.023 $\pm$ 0.013&
0.009 $\pm$ 0.013 & -0.018 $\pm$ 0.007&0.029 $\pm$ 0.013\\\hline\hline
\end{tabular}\end{center}
\vspace{0.1cm}Notes: For the samples in column 1, columns 2 to 5 give the slopes $\gamma$ between $(g-z)$
and the respective magnitude of the blue and red GC
subpopulations, as derived from \kmm determined peak positions. 
Errors come
from random resampling of the data points using their measured
dispersion around the fit. For the two simulated sets of CMDs, 
``CM'' refers to an implemented slope $\gamma_z=-$0.040
in the blue peak and no slope in the red peak. ``NoCM'' refers to no
implemented slope. Errors quoted for the simulations are rms around the mean
for 20 simulated M87 CMDs. 
The faint limiting magnitude for fitting
was $-7.7$ mag in $M_z$ and $-6.7$ mag in $M_g$. The separation
between inner and outer region for the lower four samples was defined
at a galactocentric distance of 65$''$. By that, the inner and outer
sample were of approximately equal size.  \normalsize
\end{table*}
\begin{table*}
\caption{Color-magnitude trends for red and blue globular clusters in
 ACS Virgo Cluster Survey galaxies, determined with biweight fits
\label{Fitresultsnaive}}
\small
\begin{center}
\begin{tabular}{|l|rrrr|}
\hline\hline Sample &  $\gamma_{z,\rm blue}$ &  $\gamma_{z,\rm red}$ &  
$\gamma_{g,\rm blue}$ & $\gamma_{g,\rm red}$\\\hline M49 ($M_B=-21.7$)&   -0.004 $\pm$
0.002&0.003 $\pm$ 0.003& -0.005 $\pm$ 0.006&0.010 $\pm$ 0.004\\ M87
($M_B=-21.4$)&   -0.031 $\pm$ 0.007&0.013 $\pm$ 0.005& -0.031 $\pm$
0.013&0.017 $\pm$ 0.006\\ M60 ($M_B=-21.2$)&   -0.024 $\pm$
0.006&0.001 $\pm$ 0.005& -0.027 $\pm$ 0.007&0.017 $\pm$ 0.002\\\hline
Simulated M87 NoCM&0.008 $\pm$ 0.007&-0.001 $\pm$ 0.005&0.008 $\pm$
0.004&0.009 $\pm$ 0.007\\ Simulated M87 CM&-0.029 $\pm$ 0.006&0.013
$\pm$ 0.004&-0.026 $\pm$ 0.007&0.022 $\pm$ 0.003\\\hline Group1
($-21.7<M_B<-21$)& -0.025 $\pm$ 0.002& 0.009 $\pm$ 0.003& -0.027 $\pm$
0.004& 0.021 $\pm$ 0.004\\ Group2 ($-21<M_B<-20$)& -0.025 $\pm$ 0.004&
0.017 $\pm$ 0.005& -0.023 $\pm$ 0.003& 0.025 $\pm$ 0.006\\ Group3
($-20<M_B<-18.4$)& -0.020 $\pm$ 0.002& -0.003 $\pm$ 0.009& -0.001
$\pm$ 0.005& 0.018 $\pm$ 0.010\\ Group4 ($-18.4<M_B<-15.2$)& -0.019
$\pm$ 0.004& 0.011 $\pm$ 0.007& -0.021 $\pm$ 0.004& 0.012 $\pm$
0.012\\ Group1, inner regions&-0.035 $\pm$ 0.005&0.012 $\pm$ 0.004&
-0.043 $\pm$ 0.006 & 0.022 $\pm$ 0.010\\ Group1, outer regions&-0.022
$\pm$ 0.004&0.003 $\pm$ 0.009& -0.020 $\pm$ 0.005&0.016 $\pm$ 0.004\\
Group1+2, inner regions&-0.032 $\pm$ 0.004& 0.016 $\pm$ 0.002& -0.029
$\pm$ 0.006&0.026 $\pm$ 0.004\\ Group1+2, outer regions&-0.025 $\pm$
0.005& 0.010 $\pm$ 0.006 & -0.022 $\pm$ 0.003&0.017 $\pm$
0.007\\\hline\hline
\end{tabular}\end{center}
\vspace{0.1cm}
Notes: Analogous to Table~\ref{Fitresults}. Difference is that peak
positions are now defined by the biweight of the color distribution
blue- and redwards of a luminosity independent limiting color, see
Fig.~\ref{normal}. Fig.~\ref{slopes} graphically compares the fits to
\kmm and biweight peak positions for all four luminosity groups.
\vspace{0.5cm} \normalsize
\end{table*}

\subsection{Monte Carlo Simulations to test KMM}

The difference in slopes between $z$ and $g$ raises the question as to
what the ``true'' slope actually is. To investigate this issue in more
detail, we compare in Figure~\ref{CMD3brightsim} the actual
measurements for Group 1 with two sets of 20 Monte-Carlo simulations
for the M87 CMD. The first set of CMDs has an implemented slope
$\gamma_{z}=-0.040$ for the blue GCs\footnote{Note that the slope
  $\gamma_g$ corresponding to $\gamma_z=-0.040$ is of course not
  identical to $\gamma_z$, but this difference is negligible for the
  sake of this study: $\gamma_z=-0.040 \neq
  \gamma_g=\gamma_z+\gamma_z^2=-0.0384$} and $\gamma_{z}=0$ for the
red ones; the other set has $\gamma_{z}=0$ implemented for either GC
population. These artificial CMDs were created by assuming the average
peak widths derived by \kmm for the blue and red GCs, taking into
account also the photometric errors on the colors. The luminosity
function is sampled from a kernel estimator of the original CMD of
M87, separately for each color peak.  Figure~\ref{Compsim} compares
the real and simulated M87 CMD with an implemented slope. Note an
important feature of those CMDs: the red peak is populated to brighter
absolute luminosities in $M_z$ than in $M_g$, while the blue peak does
not show such a difference. This is because, relative to GCs in the
blue population, GCs in the red population are brighter in $M_z$ than
in $M_g$.

Table~\ref{Fitresults} gives the slopes fitted to the mean colors from
\kmm for the two simulated sets of M87 CMDs, together with the slopes
for the real data.  This comparison shows that both the CMD of M87
alone, and that for the co-added CMD of Group 1, are inconsistent with
a magnitude independent color for the blue GCs. The data are
consistent with a simulated slope of --0.040 for the blue GCs and no
slope for the red GCs.

Note the positive slope found for the red GCs in both the real and
simulated data in $M_g$. Averaging over the 20 simulated CMDs, the
mean red slope of $\gamma_g \simeq +$0.03 is different from zero at
the 5.7$\sigma$ level.  That finding is noteworthy since the
simulations do {\it not} implement a slope for the red peak. This
exercise therefore demonstrates an important caveat of the \kmm
fitting: at a given magnitude, as one of the two peaks becomes
de-populated (in this case the red peak in $M_g$), its \kmm position
will artificially be dragged towards the position of the more
prominent peak. As a consequence, the mean color of the blue
population will be artificially shifted towards bluer colors. This
results in a weaker slope for the blue peak in $M_g$ than in $M_z$. In
$M_z$, both peaks are populated up to comparable absolute
luminosities, and therefore the fitted slopes for both red and blue
peaks correspond to the simulated ones. That is to say, only the {\it
  weaker} slope in $M_g$ is caused by a \kmm bias. \kmm does not
artificially introduce the very significant negative slope in the blue
$M_z$ peak: this is clearly a real trend that is intrinsic to the
data.

When interpreting the slope as a mass-metallicity relation for GCs in
the blue peak, we derive a scaling $Z\propto M^{0.48 \pm 0.08}$ by
combining the slope $\gamma_z = $-0.037 found in the previous section 
with
$$\frac{d{\rm [Fe/H]}}{d(g-z)} = 5.14~{\rm dex~mag^{-1}}$$
from Peng
\etal~\cite{Peng05} and assuming a constant mass-to-light ratio over
the luminosity range investigated. This is in good agreement with the
scaling relation $Z\propto M^{0.55}$ found by Harris
\etal~\cite{Harris06}.

\begin{figure*}
\plottwo{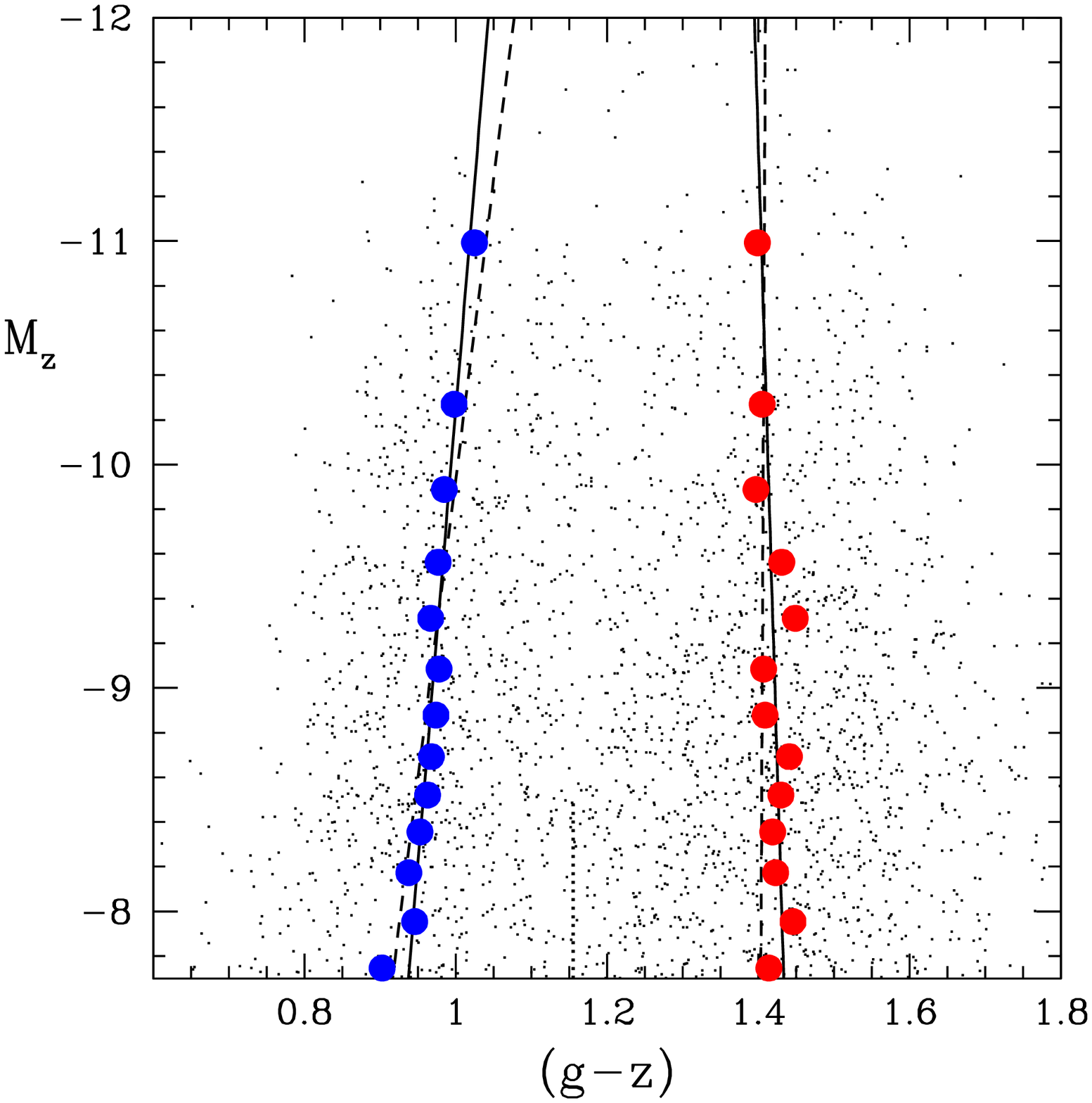}{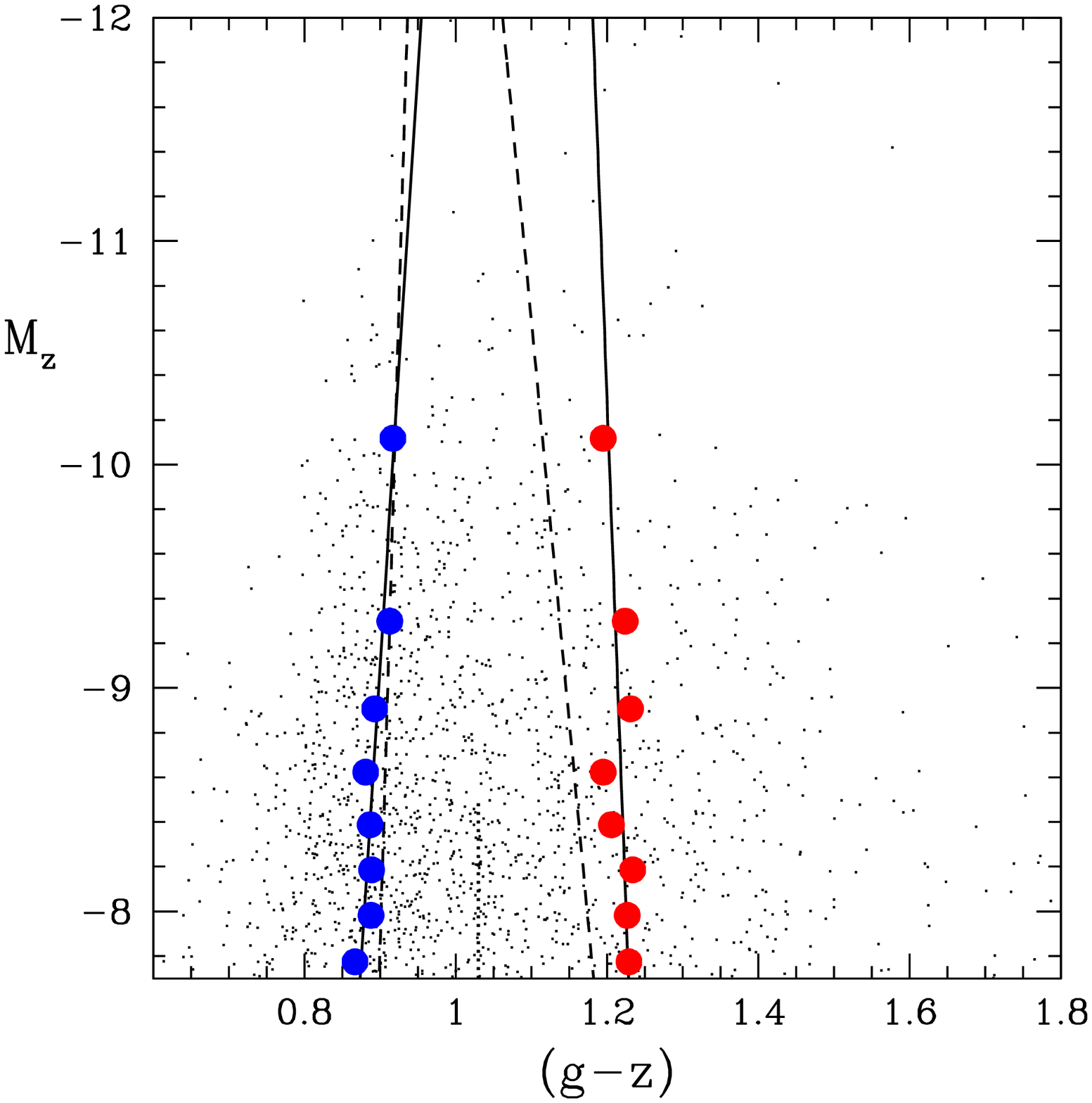}
\caption{\label{normal}{\it (Left Panel)} Color-magnitude diagram in $M_z$ for GCs
  in Group 1. Filled circles indicate the biweight estimates for the
  mean GC color bluewards and redwards of the dividing color
  $(g-z)=1.155$, which is indicated as a dotted vertical tick mark.
  The solid lines are least-squares fits to the circles. For
  comparison, the dashed lines give the linear fit obtained from the
  \kmm colors (Table~\ref{Fitresults}). {\it (Right Panel)} Same as in
  the left panel, but for Group 4. The offset between the \kmm and
  biweight fits for the red subpopulation is due to its asymmetric
  color distribution (i.e., an extended tail towards red colors).}
\end{figure*}

As an alternative to {\tt KMM}, we have calculated the biweight
location (Beers \etal~1990) for the GC subpopulations
blue-wards and red-wards of a magnitude independent limiting color.
This limiting
color is the mean of the \kmm position of the blue and red populations,
weighted by the average width of either peak. The results for this
alternative method are shown in Table~\ref{Fitresultsnaive}. As an
example, we show in Fig.~\ref{normal} the peak positions for
Groups 1 and 4 in $M_z$. Generally speaking, the errors for this
biweight method decrease significantly compared to the \kmm data
points, and the results confirm the significant slope found for the
blue GC population. They also confirm that no significant
correlation is found for M49. Table~\ref{Fitresultsnaive} and
Figure~\ref{slopes} show that the difference between fits in $M_g$ 
and $M_z$ is considerably smaller than in the case of {\tt KMM}.

Nevertheless, this alternative method is itself not immune
to biases, as is clear when it is applied it to the simulated CMDs (see
Table~\ref{Fitresultsnaive} and Figure~\ref{slopes}). In the case of the
blue population exhibiting a negative slope, a fixed limiting magnitude
between both populations will systematically decrease the absolute slope
of the blue population, and introduce an artificial positive slope
for the red.

\section{Environmental Effects}
\label{environment}

Having examined how specific features of the CMD may influence the
fitted color-magnitude relations for the blue and red GC populations,
we now investigate the extent to which possible environmental effects
may influence the observed trends.

\subsection{Dependence on Host Galaxy Luminosity}

\begin{figure}
  \centerline{\plottwo{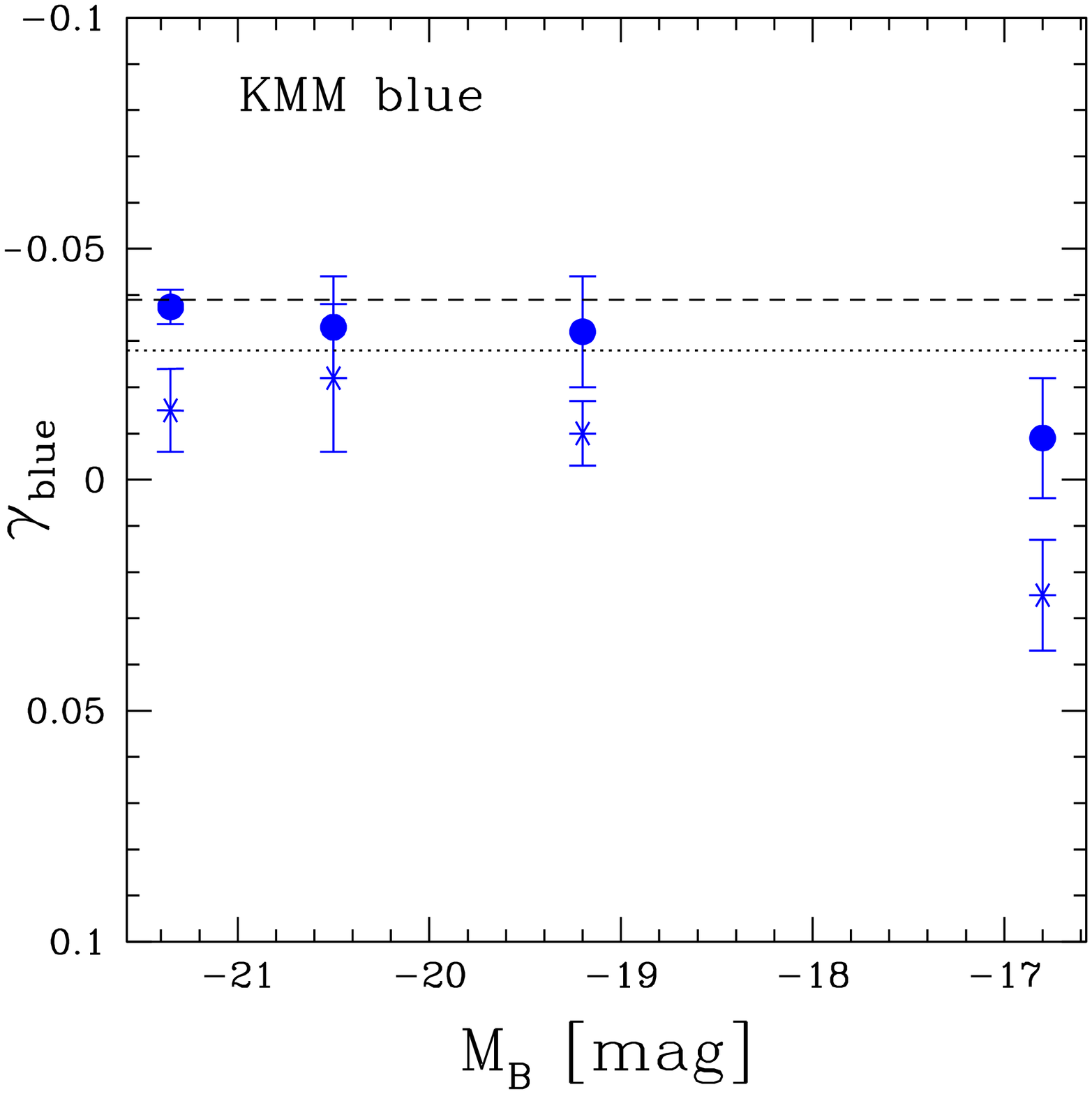}{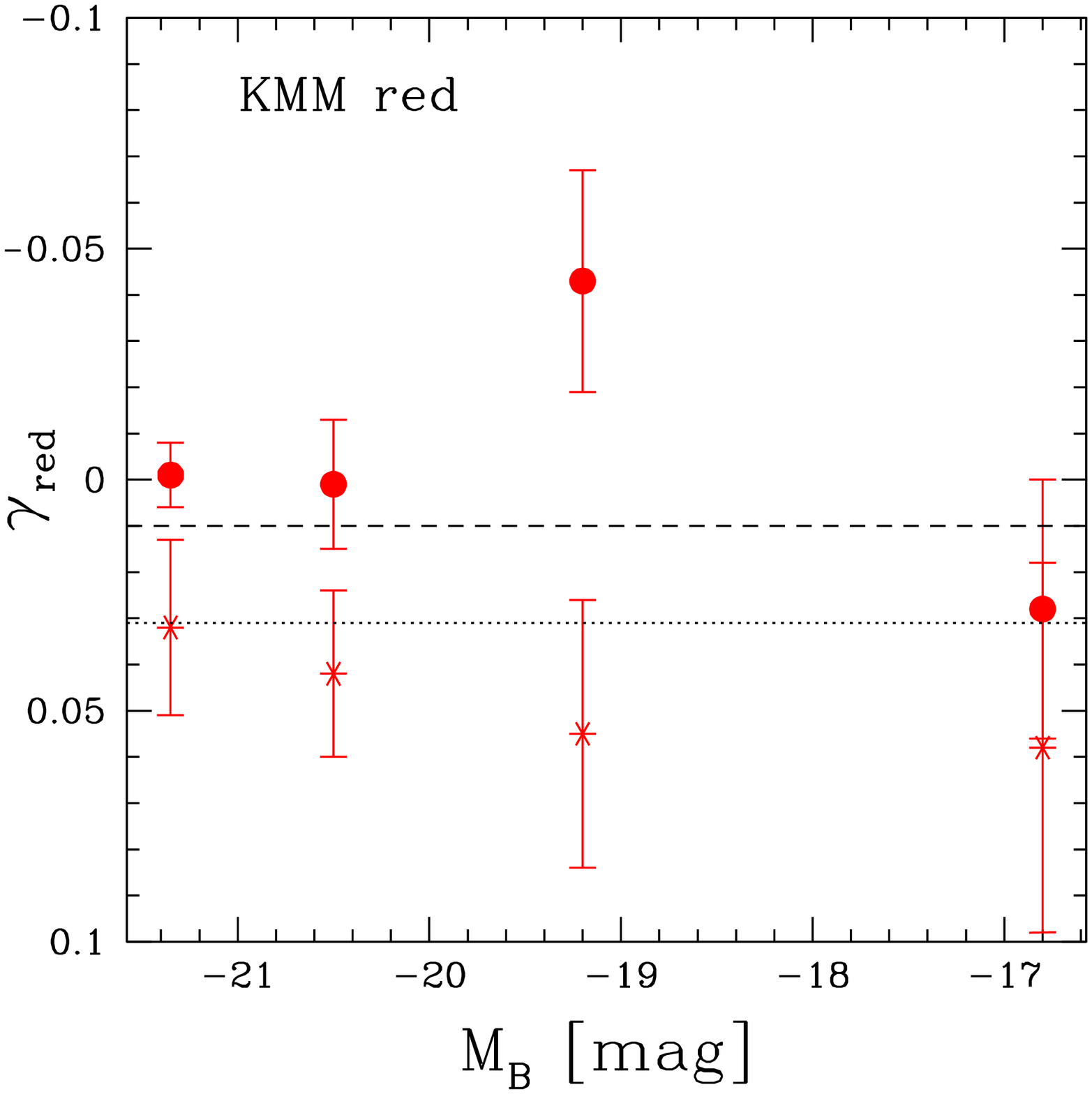}}
  \centerline{\plottwo{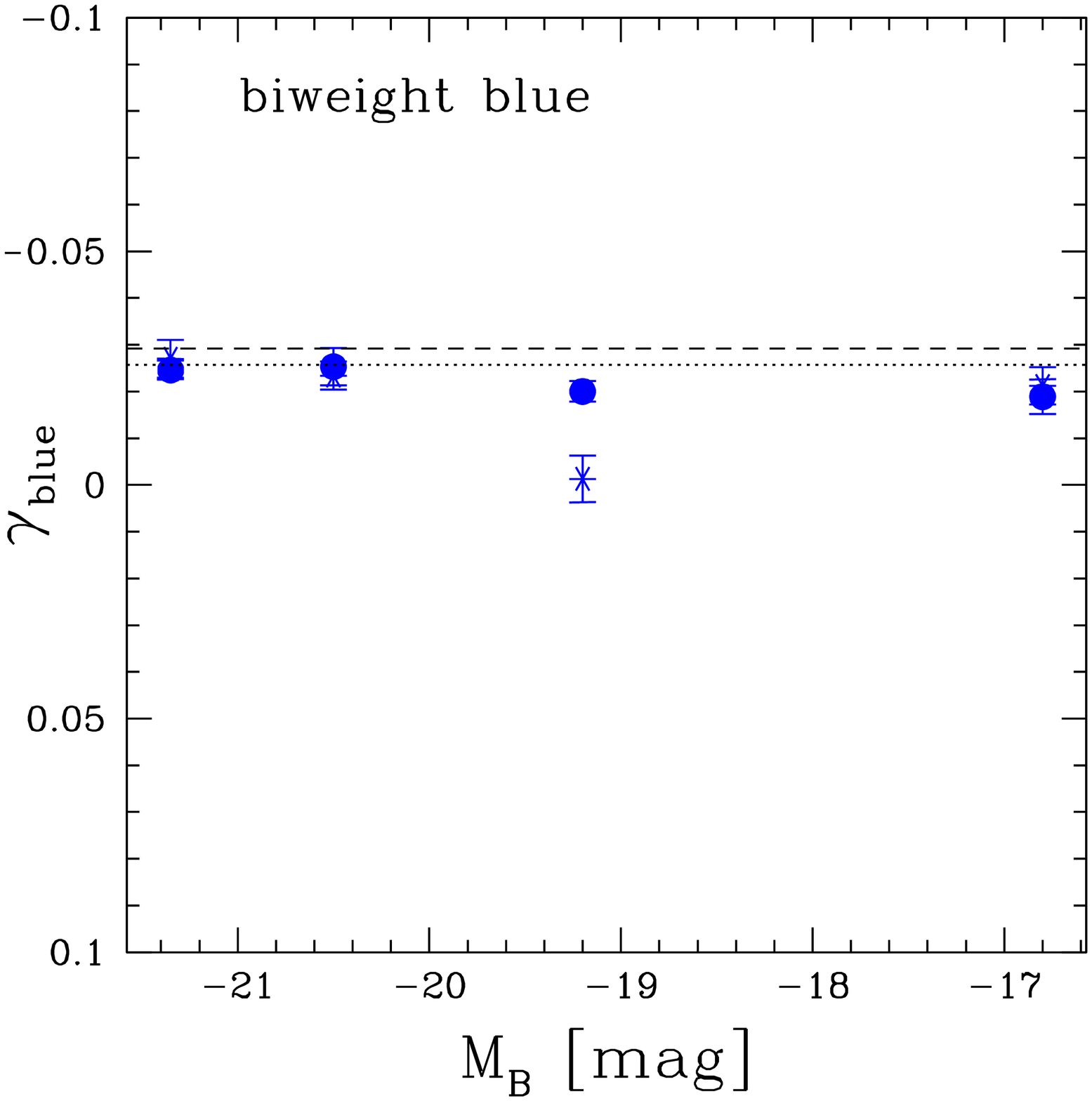}{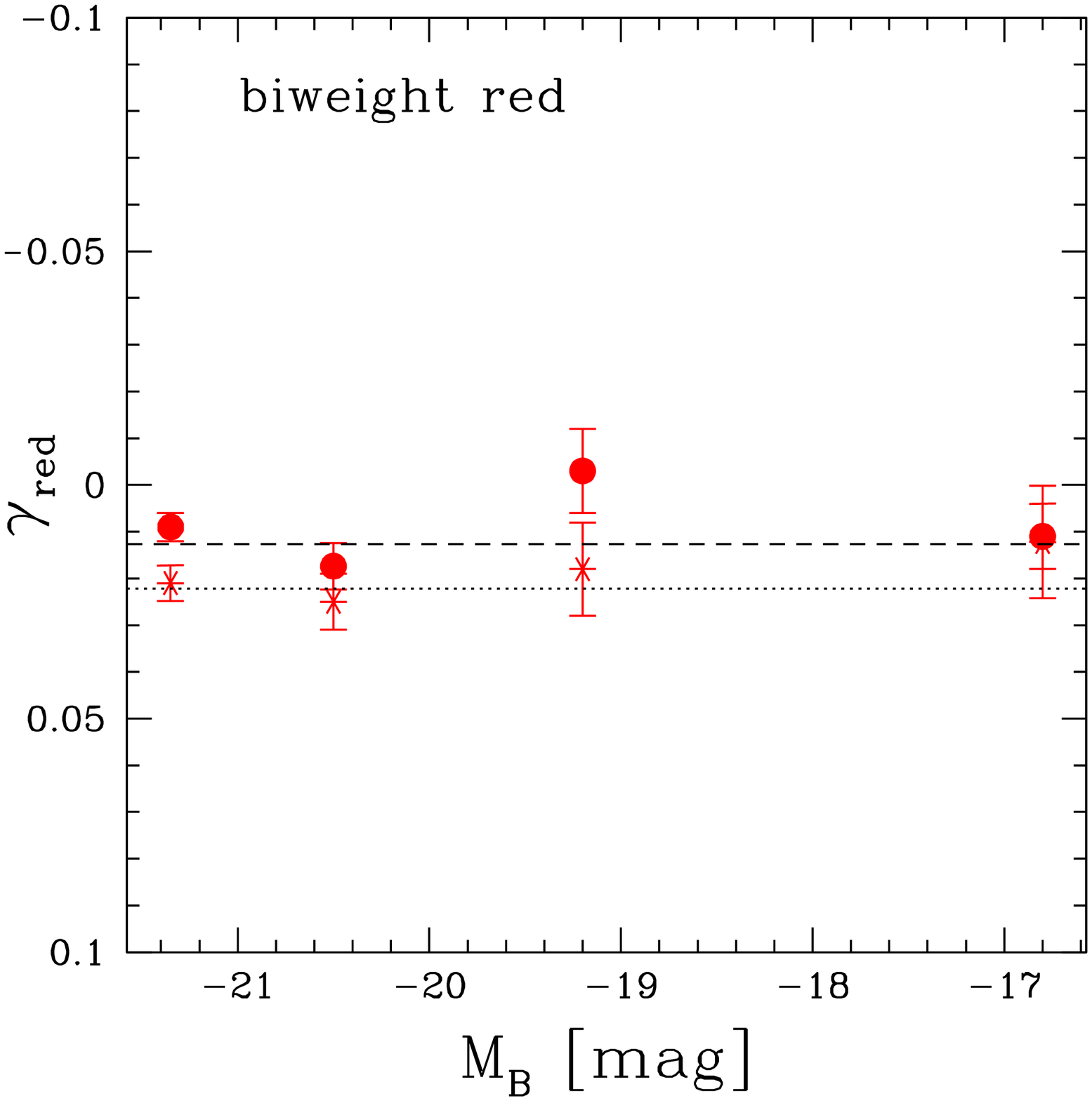}}\caption{\label{slopes}
    Slopes $\gamma$ of the GC color-magnitude relation for the four
    host-galaxy luminosity groups plotted against the respective mean
    host galaxy magnitude. Filled circles and asterisks show results
    for $\gamma_z$ and $\gamma_g$, respectively. {\it (Upper Panels)}
    \kmm fits (see Table~\ref{Fitresults}).  {\it (Lower Panels):}
    Biweight fits (see Table~\ref{Fitresultsnaive}).  {\it (Left
      Panels)} Slopes for the blue subpopulations.  {\it (Right
      Panels)} Slopes for the red subpopulations. The horizontal lines
    indicate the respective fit results to the artificial CMD of M87,
    with a slope of $\gamma_z=-0.04$ implemented for the blue
    subpopulations and zero slope for the red subpopulations. Dashed
    and dotted lines correspond to $\gamma_z$ and $\gamma_g$,
    respectively.}
\end{figure}

Figure~\ref{CMDclasses} shows CMDs for Groups 1--4
with the \kmm peaks over-plotted. Derived slopes for the color-magnitude
relations in each group are presented in 
Tables~\ref{Fitresults}~and~\ref{Fitresultsnaive}. The dependence of 
slope on host galaxy magnitude is illustrated in Figure~\ref{slopes}.

The \kmm fits in $M_z$ show that the slope of the blue GCs in Groups 2
and 3 are indistinguishable from that for Group 1; in all three
groups, the slope is inconsistent with zero at about the 3$\sigma$
level. This indicates that the color-magnitude trend among metal-poor
GCs is {\it not} restricted to just the brightest galaxies. Only for
Group 4 does the measured slope of the blue population become
consistent with zero. The biweight fits in the lower panels of
Figure~\ref{slopes} paint a similar picture: the slope decreases only
gradually from Group 1 to 4, and apparently does not drop to zero.
Given that the biweight fits systematically underestimate the blue
slope, this is an indication that the color magnitude trend persists
over the entire magnitude range of the ACSVCS galaxies.

The trend for the red GCs determined using \kmm becomes progressively
more uncertain for fainter galaxy magnitudes. This is because of the
decreasing fraction of red (metal-rich) GCs with decreasing host
galaxy luminosity (Peng \etal~2006). The firmest result from
Figure~\ref{slopes} regarding the color-magnitude relation for red GCs
is that there is no strong $M_z$ slope for the two brightest groups;
the slopes for the two faintest galaxy groups have large
uncertainties.  The large positive slopes found in $M_g$ are
particularly prone to \kmm biases, a conclusion that is confirmed by
looking at the more robust biweight fit results. The comparison
between the biweight fits to the real data and the simulations shows
that the slope of the red population is consistent with zero over the
full luminosity range.
\begin{figure*}
\centerline{\plottwo{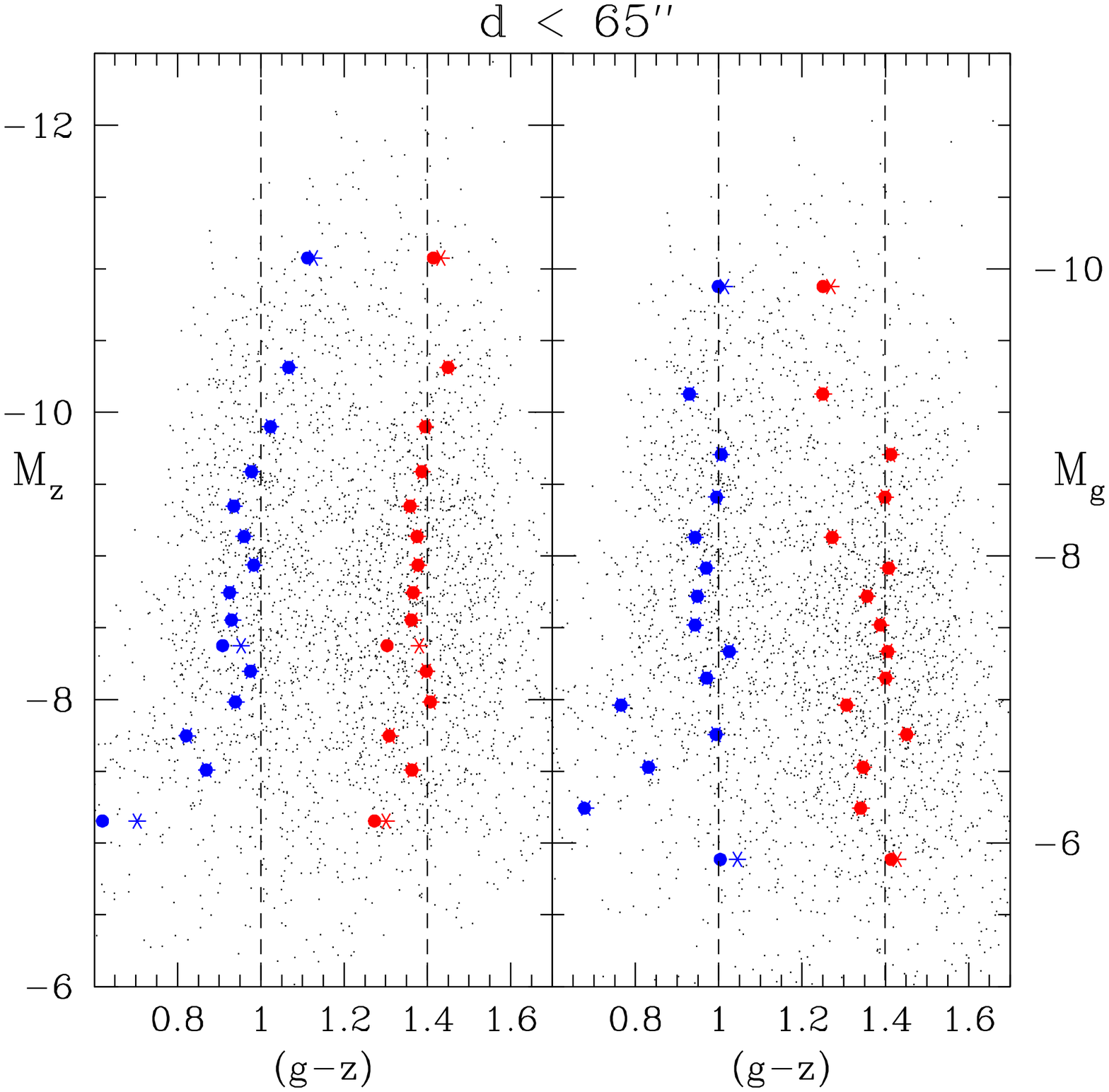}{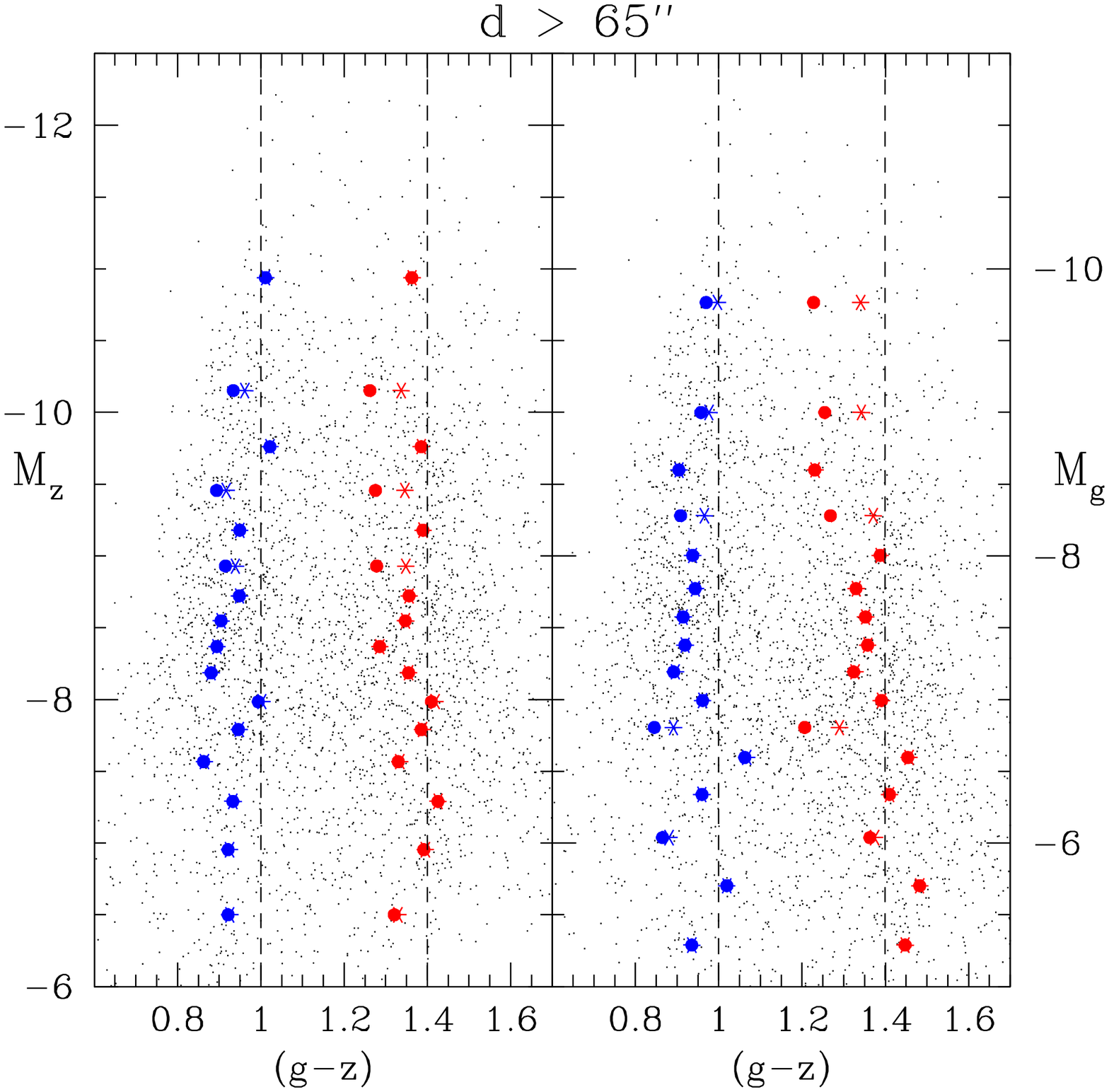}}
\caption{\label{radial}Radial dependence of the color-magnitude for the
  combined GC systems of Groups 1 and 2. {\it (Left Panel)} GCs having
  galactocentric distances $R < 65\arcsec$ (corresponding to $d
  \lesssim 5.2$~kpc).  {\it (Right Panel)} GCs having galactocentric
  distance $R > 65\arcsec$. Note the stronger trend for the blue
  subpopulation inside $65\arcsec$ (see also Table~\ref{Fitresults}).
}
\end{figure*}

\begin{figure}
\plotone{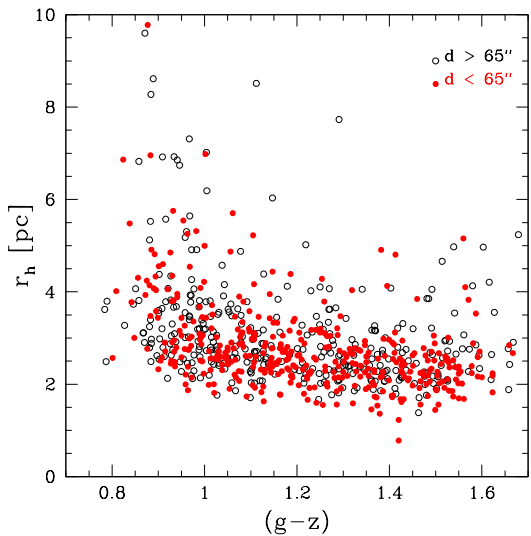}
\caption{\label{sizecolor}Half-light radius, $r_h$, plotted against $(g-z)$ 
  for GCs belonging to Group 1 and 2, and with magnitudes in the range
  $-11.5<M_z<-10$ mag (see also Figure~\ref{radial}). Open circles are
  GCs with galactocentric distances $R > 65\arcsec$, filled circles
  are GCs with $R < 65\arcsec$.  The mean size of the outer GCs is
  larger than that of the inner GCs by about 10\% (see also Jord\'an
  \etal 2005). The cumulative $r_h$ distributions of inner and outer
  GCs are inconsistent with each other at the 99.99\% confidence level
  according to a KS test.}
\end{figure}

\begin{figure*}
\plotone{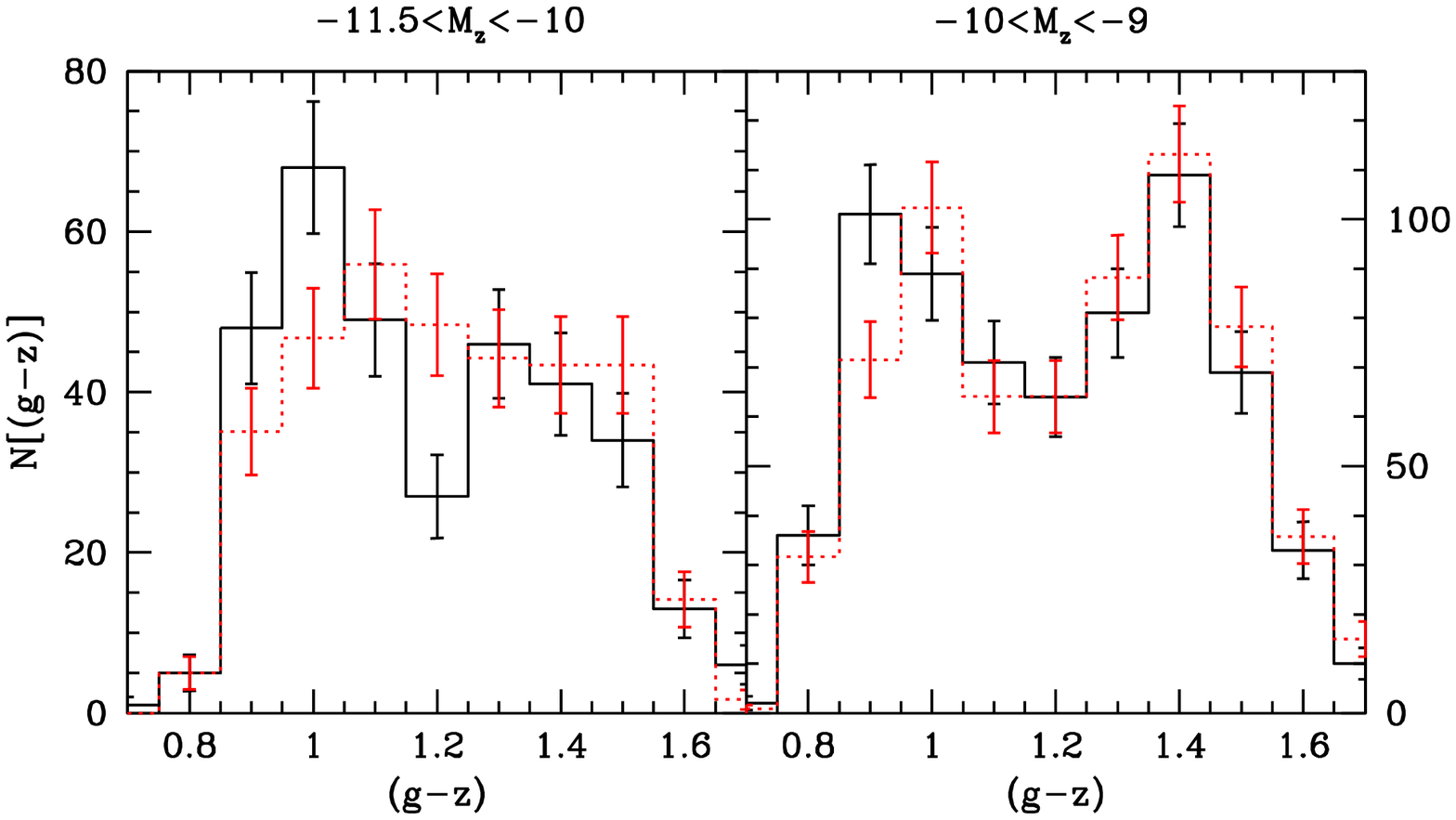}
\caption{\label{colorhist}{\it (Left Panel)} GC color distributions 
  for the two samples shown in Figure~\ref{sizecolor}. The dotted
  histogram is the sample of inner GCs with $-11.5<M_z<-10$ mag; the
  solid histogram is the sample of outer GCs with $-11.5<M_z<-10$ mag.
  {\it (Right Panel)} Same as in left panel, except for GCs with
  $-10<M_z<-9$ mag. The KS probability for the inner and outer GC
  samples having the same underlying distribution is 1\% in the left
  panel and 14\% in the right one. To facilitate the comparison, the
  histograms for the inner GCs have been rescaled by a factor of
  1.2$\times$ in both plots, so that inner and outer sample have equal
  sizes.}
\end{figure*}

\subsection{Dependence on Galactocentric Distance}

An interesting feature of the observed color-magnitude relations is
illustrated in Figure~\ref{radial}. Here we subdivide the GC samples
of Groups 1 and 2 by galactocentric radius into equally-sized,
``inner" and ``outer" subsamples. As the dividing radius, we choose $R
= 65\arcsec$, which translates to a physical distance of $\approx
5.2$~kpc at the distance of the Virgo Cluster. In terms of the
effective radii, $R_e$, of the host galaxies in Groups 1 and 2, this
fixed dividing radius corresponds to $\approx$ 0.3--0.6~$R_e$
(Ferrarese \etal 2006).
Tables~\ref{Fitresults}~and~\ref{Fitresultsnaive} report the slopes
for the different subsamples. For Group 1, the slope found for the
inner GCs is a factor of $\approx$ 2--3 higher than that for the outer
GCs.  Both the \kmm and biweight methods support this conclusion, at
significance levels of 2.2$\sigma$ and 2.5$\sigma$, respectively.  For
the combined Group 1+2 samples, the \kmm fits still obtain a
difference of about a factor of three (at 2.9$\sigma$ significance),
while the differences are almost insignificant for the biweight fits.

In Figure~\ref{sizecolor}, we plot color vs. half-light radius for GCs
with $-11.5<M_z<-10$ mag, separated into inner and outer samples.
There is a small size offset between the inner and outer sample, in
the sense that outer GCs are larger by about 10\%. This dependence of
$r_h$ on galactocentric radius has already been investigated by
Jord\'an \etal \cite{Jordan05a}.  In Figure~\ref{colorhist}, we plot
the color histograms for two luminosity ranges: in the left panel, we
show GCs in the same luminosity range as in Figure~\ref{sizecolor}; in
the right panel, we show GCs in the range $-10<M_z<-9$~mag. Clearly,
the gap between the two peaks is larger among the fainter sample of
GCs, a consequence of the color-magnitude trend. Furthermore, for the
brighter GCs, the inner and outer color histograms appear to differ
more than for the fainter GCs. The KS probability with which the inner
and outer sample of the bright GCs have the same underlying
distribution is 1\%, while it is 14\% for the faint sample. Indeed,
while the bimodality feature is still notable for the outer bright
GCs, it is slightly smeared out for the bright, inner GCs. This
suggests that the dependence of the color-magnitude relation on
galactocentric distance is driven mainly by the bright GCs ($M_z<-10$
mag).

Figure~\ref{colorhist} also shows that the fraction of blue-to-red GCs 
decreases slightly for smaller galactocentric distance (see also, e.g.,
Dirsch \etal 2003). The corresponding fractions of blue-to-red 
GCs are 0.45:0.55 and 0.50:0.50 for the inner and outer GCs, respectively.
It is apparent from the color histograms in Figure~\ref{colorhist} 
that this small effect cannot be responsible for the steeper
color-magnitude relation found for the inner sample of GCs.


\subsection{Comparison to CMDs of Other Stellar Systems}
\label{comparison}

Before proceeding, we pause to compare the distribution of GCs in
the CMD with those of other stellar systems: i.e., the nuclei of
ACSVCS galaxies, the ACSVCS galaxies themselves, and some
candidate ultra-compact dwarf galaxies (UCDs) identified in the
course of the ACSVCS. 

\begin{figure}
\plotone{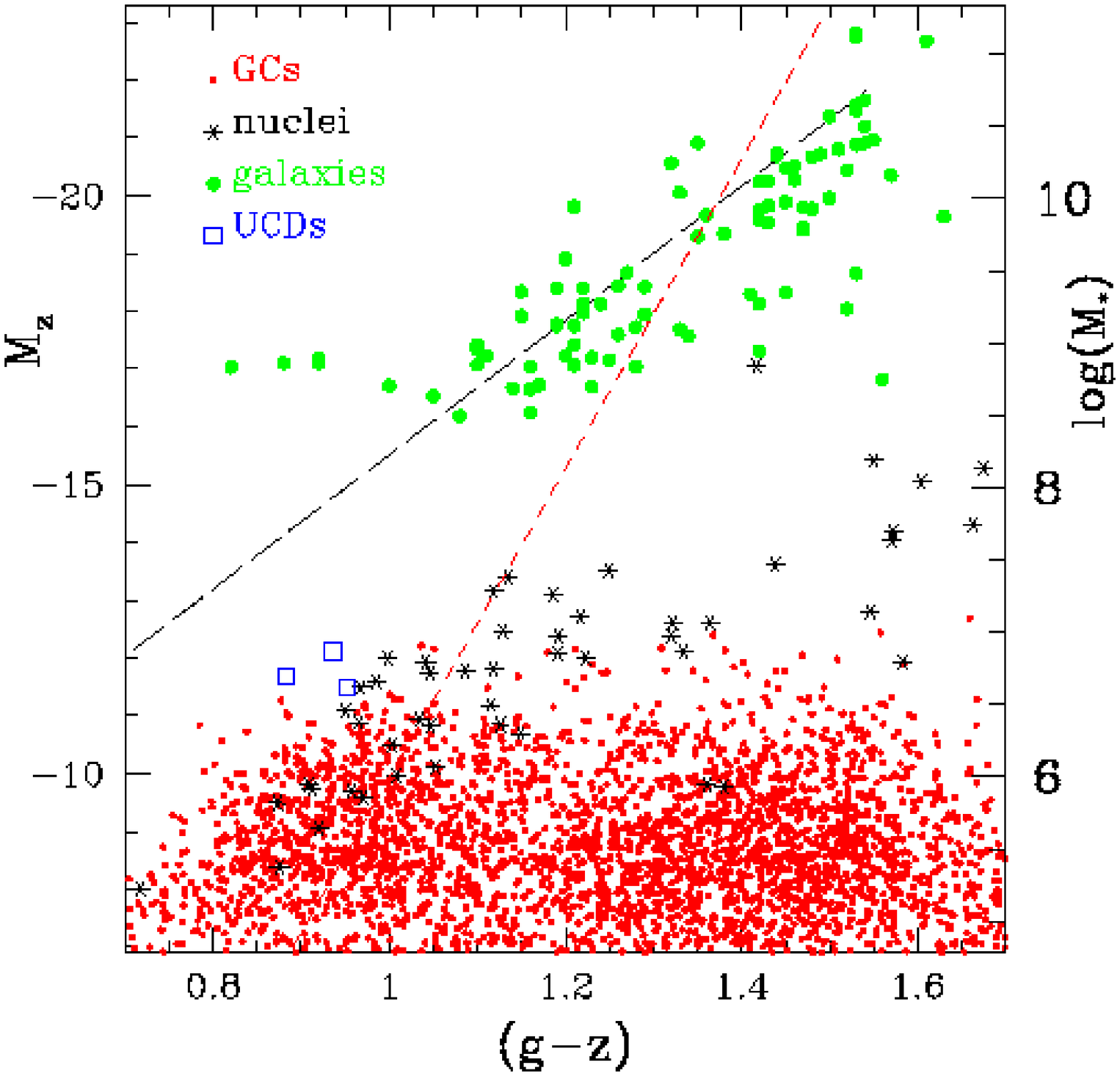}
\caption{\label{cmdallobjects}Color-magnitude diagram for GCs, compact
  stellar nuclei, early-type galaxies and UCDs in the Virgo Cluster.
  Small filled circles (online red) are GCs belonging to Group 1. The
  short dashed line is the \kmm fit to the blue GC subpopulation,
  extrapolated to brighter luminosities.  Asterisks are the nuclei of
  ACSVCS galaxies from C\^ot\'e \etal~\cite{Cote06}.  Open squares
  (online blue) are the three UCD candidates from Ha\c{s}egan
  \etal~\cite{Hasega05} with ACS imaging.  Large filled circles
  (online green) are the ACSVCS galaxies themselves (Ferrarese
  \etal~2006). The long dashed line indicates the $(V-I)$
  color-magnitude relation for early-type galaxies in the Fornax
  Cluster from Hilker \etal~\cite{Hilker03}.  The Fornax relation was
  transformed into $(g-z)$ according to $(g-z)=2.64(V-I) - 1.72$,
  following Fukugita \etal~\cite{Fukugi96}. The ordinate on the right
  denotes stellar masses, converted from $M_z$ using the mean
  M$_*$/L$_z$ ratio at metallicities of [Fe/H] = --2.25 and +0.56~dex
  for a 10-Gyr population (Bruzual \& Charlot~2003), assuming a
  Chabrier~\cite{Chabri03} IMF.  Due to the low sensitivity of $M_z$
  on metallicity, M$_*$/L$_z$ varies by just 40\% between the two
  extreme metallicities, justifying a color independent
  luminosity-mass transformation for illustrative purposes. }
\end{figure}

Figure~\ref{cmdallobjects} compares the combined CMD of GCs (red
circles), compact stellar nuclei (asterisks; C\^{o}t\'{e} \etal~2006),
ACSVCS galaxies (green circles; Ferrarese \etal~2006) and UCDs (open
squares; Ha\c{s}egan \etal 2005). In this figure, we have converted
from $M_z$ to stellar masses using the population synthesis models of
Bruzual \& Charlot~\cite{Bruzua03} under the assumption of a 10-Gyr
old population and a Chabrier IMF (Chabrier~2003).  For comparison, we
also show the color-magnitude relation for early-type galaxies in the
Fornax Cluster from Hilker \etal~\cite{Hilker03}, after converting
from $(V-I)$ to $(g-z)$ using the transformation of Fukugita
\etal~\cite{Fukugi96}.

The majority of the blue GCs are not consistent with an extrapolation of
the galaxy color-magnitude relation, nor with that of the nuclei. Moreover,
the color-magnitude relations of the nuclei and the galaxies are steeper
than that of the GCs. Only the faintest nuclei appear to be roughly
consistent with the color-magnitude relation observed for the blue GCs.
In any case, we point out that the existence of a color-magnitude trend
for the blue GCs results in a paucity of GCs at high luminosities and
blue colors: i.e., $M_z \lesssim -10$ mag and $(g-z) \lesssim 0.9$ mag. It is 
difficult to understand how contamination of the observed CMDs by some
unrecognized stellar system --- stripped nuclei, ultra-compact
dwarfs (UCDs), or otherwise --- can lead to an underpopulation of this region
of the CMD.

\section{Discussion}

As a starting point for this discussion, we summarize the two main
findings thus far, both of which concern dependencies of the measured
color-magnitude slope on external factors: (1) The slope for the blue
GCs decreases gradually towards fainter host galaxy luminosities, whereas
there is no significant color-magnitude relation for the red GCs; and 
(2) the slope is steeper for GCs which lie closer to the centers of 
their host galaxies.

We now discuss a number of scenarios that that might help explain the
observed trend: (1) contamination by super star clusters, stripped
nuclei, or UCDs; (2) accretion of GCs from low-mass galaxies; (3)
stochastic effects; (4) capture of field stars by individual GCs; and
(5) GC self enrichment. In \S\ref{metallicity}, we briefly discuss how
the observed color-magnitude trends appear after transforming into the
metallicity-mass plane.

\label{discussion}
\subsection{Super Star Clusters, Stripped Nuclei, and UCDs}
\label{ssc}

One idea that may explain the slope difference between inner and outer
GCs is the presence of overluminous star clusters of intermediate
color that may lie close to the galaxy centers. Such a putative
population would populate the region between the two color sequences
at high luminosities.  Merged super star clusters (e.g., Fellhauer \&
Kroupa~2002, 2005; Kissler-Patig \etal~2006) could, in principle, be
those objects. They are expected to be more extended than their
monolithic counterparts of equal mass (e.g., Bekki \etal~2004;
Fellhauer \& Kroupa~2002; Bastian \etal~2005).  However, in the size
vs. color plot shown in Figure~\ref{sizecolor} it is the {\it outer}
GCs that are slightly larger. More importantly, there is no offset in
the size distribution for the inner GCs with intermediate colors
(i.e., $\approx (g-z)=1.20$~mag). This makes the presence of a large
number of merged star clusters unlikely.

Another possibility is the presence of ``naked" or tidally stripped
dE,N nuclei, as suggested by Harris \etal~\cite{Harris06} (see also
Bassino {\it et~al.} 1994).  The dE,N nuclei of Virgo Cluster galaxies
do exhibit a color-magnitude trend and have colors comparable to those
of the blue GCs (see, e.g., Lotz \etal~2004; C\^ot\'e \etal~2006;
Figure~\ref{cmdallobjects}). C\^{o}t\'{e} \etal~\cite{Cote06} find a
median half-light radius of $r_h=3.2\pm0.6$~pc for nuclei with
$-11.5<M_z<-10$~mag. This is comparable to the sizes of GCs in this
same luminosity range (Figure~\ref{sizecolor}).  Stripped nuclei might
therefore be confused with ``normal" blue GCs at the highest
luminosities.

However, as pointed out in \S\ref{comparison}, the overall
distribution of GCs with the CMD shows a {\it lack} of bright, blue
objects. Very generally, this dearth is not explained by the addition
of an ``extra" population of contaminators, but rather requires an
underlying trend for most of the objects. Stripped nuclei would have
to dominate the population of blue GCs.  This seems unlikely to us for
several reasons. First, the nuclei luminosity function peaks $\approx$
2--4~mag brighter than that of GCs (see Lotz \etal~2004;, C\^{o}t\'{e}
\etal~2006, see also Figure~\ref{cmdallobjects}). More seriously,
though, is the shear number of nuclei that would be required in this
scenario.  Observationally, there is an upper limit on the number of
naked nuclei brighter than $M_V \approx -11$~mag from Jones {\it
  et~al.} (2006). These authors found nine compact objects -- UCDs in
their notation -- with $M_B<-10.7$ mag in a survey that is almost
complete within the central degree of the Virgo Cluster (excluding the
1-2\arcmin~surrounding M87). Their nine compact objects are
distributed over an area that is $\sim$ 100$\times$ the ACS field of
view, in qualitative agreement with the predictions by Bekki {\it et
  al.}~\cite{Bekki03a} for the spatial distribution of tidally
stripped dE,N nuclei. Therefore, a presence of dozens, or possibly
hundreds, of naked nuclei in our ACS fields for the most luminous
galaxies seems unlikely.

A final note regarding the possibility of UCDs as contaminators: the
two most discussed formation channels for these compact stellar
systems are stripped dE,N nuclei and merged stellar super-clusters
(e.g. Minniti {\it et al.}~1998, Hilker \etal~1999a, Drinkwater {\it
  et al.}~2003, Ha\c{s}egan \etal~2005, Kissler-Patig \etal~2006,
Mieske \etal~2006). Since these two formation channels correspond to
the two kinds of contaminators discussed in this Section, we conclude
that also UCDs are unlikely to account for the observed trends.

\begin{figure}
\plotone{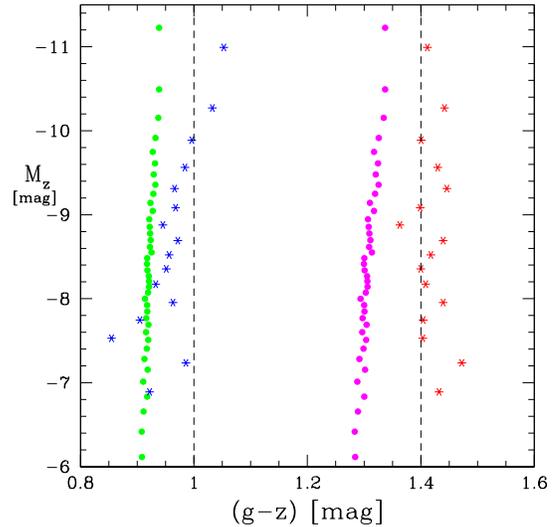}
\caption{\label{cmdall}One test of the accretion scenario, showing a CMD 
  in $M_z$ and $(g-z)$. Asterisks indicate the \kmm fits to the blue
  and red subpopulations for GCs in Group 1 (see
  Figure~\ref{CMDclasses}). Dots show the mean GC color as expected
  from the mean host galaxy luminosity in each luminosity bin for the
  co-added CMDs of Groups 2-4. The tilt seen for the dots is the
  result of two effects: (1) the larger width of the GC luminosity
  function at brighter host galaxy magnitudes; and (2) the correlation
  between the mean GC color and host galaxy luminosity. See text for
  details.}
\end{figure}

\subsection{Accretion from Low-Mass Galaxies}
\label{accretion}

A population of GCs accreted from galaxies with a range of masses
(e.g. C\^{o}t\'{e} \etal~1998 or Hilker \etal~1999b) will exhibit a
color-magnitude relation if: (1) the mean GC color of a single galaxy
scales with galaxy mass; and (2) more massive GCs preferentially form
in more massive galaxies.  Indeed, both conditions are satisfied in
actual early-type galaxies (see, e.g., Peng \etal~2005; Kundu \&
Whitmore~2001; Jord\'an {\it et~al.} 2006 in prep. and references
therein). Moreover, this simple scenario may offer, at least in
principal, a way to qualitatively understand the slope differences for
individual galaxies (e.g., the non-existence of a correlation for M49)
as a consequence of different accretion/merger histories.

In order to quantify the tilt in the CMD created by the accretion of
GCs from low-mass galaxies that do not themselves have color-magnitude
relations, we consider the combined GC sample of Groups 2--4 (see
Figure~\ref{cmdall}).  We first calculate the mean host galaxy
magnitude for each bin, and then calculate the color in each GC bin
according to the relation between host galaxy magnitude and mean color
of the blue and red GC subpopulations (Table~3 of Peng \etal~2006).
The outcome of this exercise is shown in Figure~\ref{cmdall}. As
expected, there is a (weak) trend between mean color with luminosity
for the GC subpopulations.

This trend is compared to the \kmm fits for Group 1. There is a
substantial offset in the sense that the fainter luminosity groups
have bluer GC colors than Group 1. In addition, the simulated slope is
substantially smaller than measured for the ensemble of galaxies in
Group 1.  All in all, we conclude that the observed color-magnitude
relations are unlikely to have been created by accretion of GCs from
low-mass galaxies, although we cannot rule out the possibility that
accretion has contributed to the observed trend.

\subsection{Stochastic effects}
\label{stochastic}

The light of old GCs is in the red optical to near-infrared pass-bands
dominated by red giant stars. Generally, the number of red giants is
only a small fraction of the total number of stars in a given stellar
population. It is therefore worth discussing to which extent
stochastic effects due to small numbers of red giants can skew the
probability distribution of the integrated GC luminosity.  This may
cause a trend of GC color with integrated luminosity, because any
stochastic effect is weaker for higher GC luminosity (larger total
number of stars) and for bluer bands (lower average stellar
luminosity).

In the works by Cervi\~{n}o \etal~(2002), and Cervi\~{n}o \& Luridiana
(2004; 2006), the effect of discrete isochrone sampling on integrated
photometric properties is discussed in detail.
  %
In Cervi\~{n}o \& Luridiana~(2006), the width, skewness and kurtosis
for the probability distribution of integrated luminosity of a stellar
population are calculated as a function of star number and age. The
faint magnitude limit of our GCs ($M_z \simeq -7.7$ mag) corresponds
to about 10$^5 M_{\sun}$, and hence about $2\times 10^5$ stars. In the
$I$-band -- taken to represent the $z$ band due to lack of estimates
for the latter band --, the photometric uncertainty arising from
sampling effects amounts to about 0.06 mag for such a population
(Cervi\~{n}o \& Luridiana~(2006); and Cervi\~{n}o, private
communications). The skewness of the distribution in integrated
luminosity is about 0.25.  The product of these two numbers -- about
0.015 mag --- is a measure of the difference between mode and mean of
the probability distribution in integrated luminosity.  Given that the
colour-magnitude trend in our data requires an almost ten times larger
color shift, stochastic effects do probably not have a significant
contribution to the observed trend.

  
There is also a phenomenological counterargument against the
importance of stochastic effects: the color-magnitude trend is shaped
more by brighter GCs (see Figure~\ref{colorhist}), while stochastic
effects are stronger for fainter GCs. Furthermore, the environmental
dependencies of the trend cannot be explained by purely intrinsic
stochastic effects.

\subsection{Capture of Field Stars}
\label{capture}

Given that the strength of the color-magnitude trend depends on local
environment, it is reasonable to examine the extent to which capture
of field stars may help to explain the observed slopes.  The capture
of field stars has previously been considered by Bica
\etal~\cite{Bica97} for the case of a 10$^5$M$_{\odot}$ star cluster
orbiting within the Galactic bulge at a distance of 1 kpc. They found
that such a cluster could absorb a substantial fraction of its total
mass during its lifetime. Similarly, Kroupa~\cite{Kroupa98} has shown
that stellar super-clusters created in disks like that of the Milky
Way can capture several tens of percent of their total mass in form of
disk stars.

Field stars of the brightest early-type galaxies are significantly
redder than their blue GC population, typically by about 0.5 to 0.6
mag (Peng \etal~2006). Therefore, there will be a color
shift towards the red by about 0.10-0.15 for a blue GC by the time that
$\approx$ 20-25\% of the GC light originates from field stars. This
effect will be notable mainly for the blue GC population, since the
color of red GCs is much closer to that of field stars. As will become
clear from the subsequent discussion, the capture rate per unit
GC mass is higher for more massive GCs, which will lead to a slope in
color-magnitude space. This qualitatively explains finding (1) from
above. In addition, the effect will be stronger in the inner part of
the host galaxy because of the higher field star density in such regions,
consistent with finding (2). Thus, field star capture seems
qualitatively consistent with the observed trend.

We now quantify this scenario by considering a GC with a
mass of 10$^6 M_{\odot}$, a mass that roughly corresponds to the brightest
clusters in Groups 1 and 2 in Figure~\ref{CMDclasses}.
For an assumed M/L=2.5, this corresponds to a total of 4$\times 10^5$ solar
luminosities. We assume the GC to be on a circular orbit around
M87. This is certainly an extreme assumption, given that C\^{o}t\'{e}
\etal~\cite{Cote01} rule out such tangentially biased orbits for the
GC system {\it as a whole}.  We nevertheless use it for the sake of
this simple estimate.  We adopt a radius of 3 kpc, representative for
the sample  within 5 kpc projected galactocentric distance.

We may estimate the number of captured stars by multiplying the total
number of stars within the volume of influence of the GC with the
fraction of stars having relative velocities below the GC escape
velocity. As the volume of influence, we define a torus around the
orbit of the GC with an inner radius of the GC's tidal radius with
respect to its host galaxy M87.  For calculating the tidal radius, we
use the mass profile of M87 derived by Romanowsky \&
Kochanek~\cite{Romano01}. They quote the integrated mass at a distance
of 3 kpc to be 1.1$\times$10$^{11}$ $M_{\odot}$, which is comparable
to the value found by C\^{o}t\'{e} \etal~\cite{Cote01}. Together with
the 10$^6 M_{\odot}$ mass of the GC, this gives a tidal radius of 9
pc. The total volume of the torus then is 4.8$\times 10^6$ pc$^3$. We
use the same paper by Romanowsky \& Kochanek to get the 3-dimensional
stellar density of M87 at 3 kpc, which is about 0.06 L$_{\sun}$
pc$^{-3}$. That is, along one orbit there will be a total of 2.9
$\times 10^5$ solar luminosities within the volume of influence of our
test GC.

However, only a small fraction of those will be captured: the average
escape velocity $v_{esc}$ of the GC within its tidal radius of 9pc is
about 37 km~s$^{-1}$, while the stellar velocity dispersion in M87 is about
275 km~s$^{-1}$ (Romanowsky \& Kochanek~2001). The GC population
in M87 has a velocity dispersion of $\sim$ 400 km~s$^{-1}$. The probability 
distribution for the
relative velocity between the field stars and a GC is given by the
convolution of both distributions, which is a Gaussian of width $\sigma$ = 485
km~s$^{-1}$. The fraction of stars (in solar luminosities) that can be
captured in the 1D case is given by the area below the Gaussian
between +/- $\frac{v_{esc}}{\sqrt{3}}$, which is 0.035. For the full
3D case, this is $0.035^3=4.3\times 10^{-5}$, or one out of 2.3
$\times 10^4$ stars. Given that in total 2.9 $\times 10^5$ stars are
in the volume of influence in each orbit, this implies a capture rate
of one dozen stars per orbit\footnote{Note that close encounters with GC stars 
may indeed alter the capture rate,
but an accurate assessment of this effect would require dedicated
$N$-body simulations, which is beyond the scope of this paper.}.
For a Keplerian orbit at 3 kpc, one
revolution lasts about 2.5$\times 10^7$ years. In a Hubble time, one
can therefore capture about 6000 stars.
This is 1.5\% of the total assumed GC luminosity, not sufficient to
fully explain the observed effects. 
Only when considering a more extreme orbital distance of 1 kpc
does  the total number of captured stars increase to 
about 4$\times 10^4$, i.e. $\simeq$10\% of all stars.


We now briefly examine the opposite effect of field star capture: namely
cluster star evaporation. 
As is well known, mass segregation in GCs (Gunn \& Griffin 1979;
Jord\'an~2004) will result in
evaporation affecting mainly the lowest mass stars. Lamers et
al.~\cite{Lamers06} find for clusters of solar metallicity
that evaporation will cause the overall GC color to become bluer with
time (over most of their lifetime) due to the preferential loss of red,
low-mass stars.
Could such a color change help to explain our observations?
Two effects are important: (1) the {\it maximum} blueward shift due to
evaporation is larger for more massive clusters (Lamers et
al.~\cite{Lamers06}), reaching up to $\Delta (V-I)\simeq$ 0.10~mag; 
and (2) the dissolution time, $t_{diss}$, is longer for more massive
clusters.  To qualitatively explain our observations one would require
the massive GCs to be in an early stage of evolution, so that
evaporation has not yet had a significant effect. The fainter GCs would
mainly define the color magnitude trend. However, Figure~\ref{colorhist}
shows that the slope changes between inner and outer GCs are
mainly driven by the {\it brightest} GCs. That is, evaporation is 
unlikely to account fully for the observed color-magnitude trend. 
Even more serious, evaporation cannot naturally explain 
the simultaneous presence of a trend for the blue GCs and its 
absence for the red.

\subsection{GC Self Enrichment}
\label{selfenrich}

Generally speaking, self enrichment will occur in any stellar system,
GCs included, if the system is able to retain a fraction of the gas
expelled by the first generation of supernovae and subsequently form a
new generation of stars. To create a metallicity-mass (or,
equivalently, a color-magnitude) trend via self-enrichment, one would
therefore require that more massive GCs are more efficiently
self-enriched.  The extent to which GCs in the Milky Way may have
self-enriched, however, is a controversial subject: e.g., Frank \&
Gisler~\cite{Frank76}, Smith~\cite{Smith96}, Gnedin
\etal~\cite{Gnedin02}, Parmentier \& Gilmore~\cite{Parmen01}, Dopita
\& Smith~\cite{Dopita86}, Morgan \& Lake~\cite{Morgan89}, Thoul
\etal~\cite{Thoul02}.

In this context, it is interesting to note that the most massive GC in
the Galaxy, $\omega$~Cen, is known to have a metallicity distribution
function that shows a long tail towards high metallicities (e.g.,
Norris, Freeman \& Mighell~1996; Stanford {\it et~al.} 2006).  Various
studies have found this high-metallicity population to be either
coeval, or perhaps 2--4 Gyr younger, than the cluster's primary
(metal-poor) population (e.g., Ferraro \etal~2004; Sollima et
al.~2005; Hughes \& Wallenstein~2000; Hilker \& Richtler~2000; Rey
\etal~2004, Stanford \etal~2006). This is consistent with $\omega$~Cen
having been at least partially self-enriched. Note, however, that
because of its peculiar star formation history, numerous investigators
have suggested that the cluster may be the stripped nucleus of a
now-dissolved dwarf galaxy (e.g., Freeman~1993; Hilker \&
Richtler~2000; Bekki \& Freeman~2003). If true, then it may be
incorrect to consider $\omega$~Cen to be a typical example of a self
enriched GC.

This caveat aside, if self enrichment is common to the formation of
the most massive GCs, then it may indeed help to explain our
observations. Most significantly, the lack of a correlation among the
red GC subpopulation is naturally explained in this scenario: i.e.,
the redder GCs are already metal-enriched, and a fixed amount {\it d}Z
of supernovae ejecta will cause a relatively weaker metallicity
increase, {\it d}[Fe/H], for metal-rich GCs. For instance, for a
system with [Fe/H] = $-$2 (Z = 0.0002), adding a fixed amount of
metals {\it d}Z = 0.0002 will result in {\it d}[Fe/H]=$+$0.3 dex.
According to Peng \etal~\cite{Peng05} this corresponds to a shift in
$(g-z)$ of about 0.06 mag and is about the observed amount of the
color magnitude trend.  However, for a system with [Fe/H] = $-$0.5 and
the same {\it d}Z, we only have {\it d}[Fe/H]=$+$0.013 dex, with a
corresponding color change in $(g-z)$ well below 0.01 mag.

A possible shortcoming of this self-enrichment scenario is that 
is does not explicitly account for environmental effects such as a 
dependence on galactocentric radius. On the other hand, feedback 
induced by pressure-confinement from the surrounding medium
may be able to help reconcile this scenario with observations.
\begin{figure}
\plotone{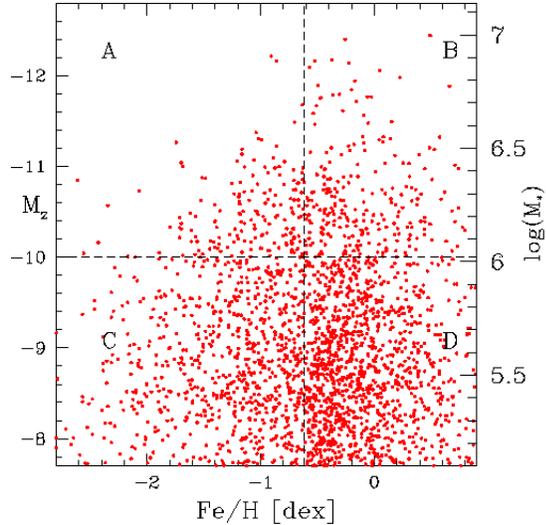}
\caption{\label{met}Metallicity plotted as a function of stellar mass and $M_z$ for 
  GCs in Group 1, assuming a quartic color-metallicity transformation
  based on the empirical calibration data of Peng \etal~\cite{Peng05}.
  Conversion from luminosity to stellar mass is performed as in
  Figure~\ref{cmdallobjects}.  The vertical dashed line corresponds to
  the limiting color between blue and red GC subpopulations from
  Figure~\ref{normal}. The horizontal dashed line corresponds to the
  dividing point between brighter and fainter GCs from
  Figure~\ref{colorhist}. A KS test shows that the metallicity
  distributions in quadrant A and C stem from the same underlying
  distribution with only 0.1\% probability. This probability is 91\%
  for the color distributions in quadrants B and D. The probability
  for the joint sample of quadrants A and B compared to the joint
  samples C and D is 16\%.}
\end{figure}

\subsection{The Distribution of GCs in Metallicity-Mass Space}
\label{metallicity}

To conclude this discussion, we show in Figure~\ref{met} a
metallicity-luminosity plot for the co-added sample of GCs in Group 1.
To estimate GC metallicities, we have applied
a quartic color-metallicity transformation based
on the empirical calibration data from Peng \etal~\cite{Peng05}. 
Note that the
resulting metallicity distribution is {\it not} bimodal, but rather shows
a single pronounced peak at [Fe/H]$\simeq$ $-$0.4~dex and a long,
extended tail towards lower metallicities. This confirms recent
warnings (e.g.,  Richtler~2005; Yoon, Yi \& Lee~2006) that bimodality 
in color does not necessarily imply bimodality in metallicity.

Using the data in Figure~\ref{met}, we can compare the metallicity
distributions of GCs brighter and fainter than $M_z=-10$ mag. For the
metal-poor population, ([Fe/H]$<$-0.6), the fainter GCs have a
significantly broader tail towards lower metallicities than do the
bright ones. A KS-test gives a 0.1\% probability for a common
underlying distribution between both samples. In the metal-rich part
of the diagram, the agreement is very good, with a KS probability of
91\%.  In other words, the empirical finding of a color-magnitude
trend for the blue GCs can hence be re-phrased in terms of
metallicity: the GC systems appear to have a skewed, but unimodal,
metallicity distribution function whose metal-poor tail becomes
increasingly extended at lower luminosities/masses.

\section{Summary and Conclusions}

In this paper we have analysed the color-magnitude relation for the GC
subpopulations in early-type galaxies belonging to the Virgo Cluster.
Our sample consists of 79 galaxies, spanning the range
$-21.7<M_B<-15.2$~mag, drawn from the ACS Virgo Cluster Survey
(C\^ot\'e \etal 2004).  Our principal findings can be summarized as
follows:

\begin{enumerate}
\item Based on \kmm fits, we find a highly significant correlation
  $\gamma_z \equiv \frac{d(g-z)}{dz} = -0.037 \pm 0.004$ for the blue
  GCs in the co-added CMD of the three brightest Virgo Cluster
  galaxies (M49, M87, M60). The sense of the correlation is such that
  brightest GCs in this subpopulation are redder than their faint
  counterparts. For M87 and M60 alone, we find comparable
  correlations; by contrast, the blue GCs in M49 do not appear to
  follow such a trend.  In no galaxy do we find a significant
  correlation between $(g-z)$ and $M_z$ for the GCs associated with
  the red subpopulation.

\item Based on \kmm fits, we find the slope $\gamma_g \equiv
  \frac{d(g-z)}{dg}$ within the blue subpopulation to be much weaker
  than $\gamma_z$. In turn, we find a mild {\it positive} correlation
  $\gamma_g$ for the red subpopulation, in the opposite sense to what
  is found for the blue subpopulation.
  
\item We test \kmm on artificial CMDs with, and without, implemented
  slopes for the blue subpopulation, and no slope for the red
  population. \kmm correctly reproduces the implemented slopes when
  using the CMD $(g-z)$ vs. $M_z$. For $(g-z)$ vs. $M_g$, \kmm gives a
  biased estimate for the color of the blue population and yields a
  positive slope for the red subpopulation, although no slope had been
  implemented. We trace this bias to the fact that in $M_g$ -- unlike
  in $M_z$ -- the luminosity function of blue GCs is populated to
  significantly brighter luminosities than for the red GCs. We
  conclude that for the real data, the strong slope $\gamma_z$ is {\it
    not} an artifact of the fitting method, while the weaker slope
  $\gamma_g$ is an artificial feature of the \kmm fit.
  
\item We derive correlations independent of \kmm by using a biweight
  estimator for GCs divided on the basis of a luminosity independent
  color.  The results confirm the significant slope found by \kmm in
  $M_z$ and reveal no significant discrepancies between the slope
  measured using $M_z$ and $M_g$.
  
\item The slope measured from the blue subpopulation decreases with
  host galaxy luminosity. However, even among the faintest galaxies
  ($-18.4<M_B<-15.2$~mag) the slope is nonzero.
  
\item The color-magnitude relation exhibited by the blue GCs is
  stronger for clusters with smaller projected galactocentric
  distances. This difference is driven mainly by GCs with $M_z<-10$
  mag.

\end{enumerate}

We examine a number of physical mechanisms that might give rise to a
color-magnitude relation with these characteristics: (1) contamination
by super-star-clusters, stripped galactic nuclei, or ultra-compact
dwarf galaxies; (2) accretion of GCs from low-mass galaxies; (3)
stochastic effects; (4) capture of galactic field stars by GCs; and
(5) self-enrichment of individual GCs.  Neither contaminants nor
accretion of low-mass galaxies appear likely to fully explain our
observations. Self-enrichment in more massive GCs may be able to
account for the observed color-magnitude relation, although more work
is needed to determine if the dependence on galactocentric radius can
be explained within the context of the self-enrichment scenario.
Although the capture of field stars {\it does} lead to a dependence on
galactocentric radius that is at least qualitatively in agreement with
our observations, this mechanism does not appear efficient enough to
explain the overall size of the observed trend.

While none of these scenarios can fully account for the full set of
observations in a straightforward manner, self-enrichment and capture
of field stars seem the most promising approaches. There is, of
course, no {\it a priori} reason that a single mechanism is
responsible for the observed color-magnitude relation, and
conceptually, the combined effects of self-enrichment and field star
capture may be able to explain the observations: i.e., field star
capture could act as a second order effect, producing environmental
variations of a global mass-metallicity relation induced by
self-enrichment.  This would alleviate the discrepancy between real
and required field star capture rate estimated in this work.
Dedicated N-body simulations going beyond these simple estimates will
be required to accurately address the efficiency of field star capture
in different environments.

\label{conclusions}
\acknowledgements We would like to thank Pavel Kroupa, Mike Fellhauer,
Holger Baumgardt, H.J.G.L.M. Lamers and Miguel Cervi\~{n}o for
fruitful discussions. We thank the anonymous referee for her/his comments.
Support for program GO-9401 was provided through
a grant from the Space Telescope Science Institute, which is operated
by the Association of Universities for Research in Astronomy, Inc.,
under NASA contract NAS5-26555.  P.C. acknowledges support provided by
NASA LTSA grant NAG5-11714.  M.J.W. acknowledges support through NSF
grant AST-0205960.  S.M.  acknowledges additional support from NASA
grant NAG5-7697 to the ACS Team.  D.M. acknowledges support provided
by NSF grants AST-0071099, AST-0206031, AST-0420920 and AST-0437519,
by NASA grant NNG04GJ48G, and by grant HST-AR-09519.01-A from STScI.
This research has made use of the NASA/IPAC Extragalactic Database
(NED) which is operated by the Jet Propulsion Laboratory, California
Institute of Technology, under contract with the National Aeronautics
and Space Administration.

{}
\newpage


\begin{thebibliography}{}
\bibitem[1992]{Ashman92}Ashman, K. M., Zepf, S. E. 1992, ApJ, 384, 50
\bibitem[1994]{Ashman94}Ashman, K. M., Bird, C. M., Zepf, S. E. 1994,
AJ, 108, 2348
\bibitem[2005]{Bastia05}Bastian, N., Emsellem, E., Kissler-Patig, M.,
Maraston, C. 2006, A\&A, 445, 471
\bibitem[Bassino {\it et~al.}(1994)]{1994ApJ...431..634B} Bassino,
  L.~P., Muzzio, J.~C., \& Rabolli, M.\ 1994, \apj, 431, 634
\bibitem[2002]{Beasle02}Beasley, M. A. {\it et~al.} 2002, MNRAS, 333, 383
\bibitem[1990]{Beers90}Beers, T. C., Flynn, K., Gebhardt, K. 1990, AJ,
100, 32
\bibitem[2003a]{Bekki03a}Bekki, K., Couch, W.J., Drinkwater, M.J.,
Shioya, Y., 2003a, MNRAS, 344, 399
\bibitem[2003b]{Bekki03b}Bekki, K., \& Freeman, K. C. 2003, MNRAS, 346, L11
\bibitem[2004]{Bekki04}Bekki, K., Couch, W.J., Drinkwater, M.J.,
Shioya, Y., 2004, ApJL, 610, 13
\bibitem[1997]{Bica97}Bica, E. {\it et~al.} 1997, ApJ, 482, 49
\bibitem[1992]{Bower92}Bower, R.G., Lucey, J.R., Ellis, R.S. 1992,
MNRAS, 254, 589
\bibitem[2003]{Bruzua03}Bruzual, G., Charlot, S. 2003, MNRAS, 344, 1000
\bibitem[2001]{Burgar01}Burgarella, D., Kissler-Patig, M., Buat,
V. 2001, AJ, 121, 2647
\bibitem[2002]{Cervin02}Cervi\~{n}o, M., Valls-Gabaud, D., Luridiana,
  V., \& Mas-Hesse, J. M. 2002, A\&A, 381, 51
\bibitem[2004]{Cervin04}Cervi\~{n}o, M., Luridiana, V. 2004, A\&A, 413, 145
\bibitem[2006]{Cervin06}Cervi\~{n}o, M., Luridiana, V. 2006, A\&A, 451, 475
\bibitem[2003]{Chabri03}Chabrier, G., 2003, PASP, 115, 763
\bibitem[1998]{Cote98}C\^{o}t\'{e}, P., Marzke, R. O., West,
M. J. 1998, ApJ, 501, 554
\bibitem[2001]{Cote01}C\^{o}t\'{e}, P. {\it et~al.} 2001, ApJ, 559, 828
\bibitem[2003]{Cote03}C\^{o}t\'{e}, P., McLaughlin, D. E., Cohen,
J. G., \& Blakeslee, J. P. 2003, ApJ, 591, 850
\bibitem[2004]{Cote04}C\^{o}t\'{e}, P. {\it et~al.} 2004, ApJS, 153, 223
(Paper I)
\bibitem[2006]{Cote06}C\^{o}t\'{e}, P. {\it et~al.} 2006, ApJS,
accepted, astro-ph/0603252 (Paper VIII)
\bibitem[2003]{Dirsch03}Dirsch, B. {\it et~al.} 2003, AJ, 125, 1908
\bibitem[1986]{Dopita86}Dopita, M. A., Smith, G. H. 1986, ApJ, 304, 283
\bibitem[2003]{Drinkw03}Drinkwater, M.J., Gregg, M.D., Hilker, M. et
al., 2003, Nature, 423, 519
\bibitem[2002]{Fellha02}Fellhauer, M., Kroupa, P., 2002, MNRAS, 330,
642
\bibitem[2005]{Fellha05}Fellhauer, M.; Kroupa, P., 2005, MNRAS, 359,
223
\bibitem[2004]{Ferrar04}Ferraro, F.R. {\it et~al.} 2004, ApJ, 603, L81
\bibitem[2006]{Ferrar06}Ferrarese, L. {\it et~al.} 2006, ApJS, 164, 334
(Paper VI)
\bibitem[1993]{Freema93}Freeman, K. C. 1993, in ASP Conf. Ser. 48, 
The Globular Cluster-Galaxy Connection, ed. G. H. Smith \& 
J. P. Brodie (San Francisco: ASP), 27
\bibitem[1997]{Forbes97}Forbes, D. A., Brodie, J. P., Grillmair,
C. J. 1997, AJ, 113, 1652
\bibitem[1998]{Ford98}Ford, H. C., Bartko, F., Bely, P. Y. {\it et~al.} 1998, Proc. SPIE Vol. 3356, p. 234-248, Space Telescopes and Instruments V, Pierre Y. Bely; James B. Breckinridge; Eds.
\bibitem[1976]{Frank76}Frank, J., Gisler, G. 1976, MNRAS, 176, 533
\bibitem[1996]{Fukugi96}Fukugita, M., Ichikawa, T., Gunn, J.E. et
al. 1996, AJ, 111, 1748
\bibitem[1999]{Gebhar99}Gebhardt, K., Kissler-Patig, M. 1999, AJ, 118, 1526
\bibitem[2002]{Gnedin02}Gnedin, O. Y. {\it et~al.} 2002, ApJ, 568, 23
\bibitem[1979]{Gunn79}Gunn, J. E., Griffin, R. F. 1979, AJ, 84, 752
\bibitem[2004]{Harris04}Harris, G. L. H., Harris, W. E., Geisler,
D. 2004, AJ, 128, 723
\bibitem[2006]{Harris06}Harris, W. E. {\it et~al.} 2006, ApJ, 636, 90
\bibitem[2005]{Hasega05}Hasegan, M., Jord\'an, A., C\^{o}t\'{e}, P. et
al. 2005,  ApJ, 627, 203
\bibitem[1999a]{Hilker99a}Hilker, M., Infante, L., Vieira, G.,
Kissler-Patig, M., Richtler, T. 1999, A\&AS, 134, 75
\bibitem[1999b]{Hilker99b}Hilker, M., Infante, L., Richtler, T. 1999,
A\&AS, 138, 55
\bibitem[2000]{Hilker00}Hilker, M., \& Richtler, T. 2000, A\&A, 362, 895
\bibitem[2003]{Hilker03}Hilker, M., Mieske, S., \& Infante, L. 2003,
A\&AL, 397, L9
\bibitem[2000]{Hughes00}Hughes, J., \& Wallerstein, G. 2000, AJ, 119, 1225
\bibitem[2006]{Jones06}Jones, J. B., {\it et~al.} 2006, AJ, 131, 312
\bibitem[2004a]{Jordan04a}Jord\'an, A., Blakeslee, J. P., Peng,
E. W. {\it et~al.} 2004a, ApJS, 154, 509 (Paper II)
\bibitem[2004b]{Jordan04b}Jord\'an, A., C\^{o}t\'{e}, P., Ferrarese,
L. {\it et~al.} 2004b, ApJ, 613, 279 (Paper III)
\bibitem[2004c]{Jordan04c}Jord\'an, A. 2004, ApJ, 613, L117
\bibitem[2005]{Jordan05a}Jord\'an, A., C\^{o}t\'{e}, P., Blakeslee,
J. {\it et~al.} 2005, ApJ, 634, 1002 (Paper X)
\bibitem[2003]{Karick03}Karick, A., Drinkwater, M.J., Gregg,
M.D. 2003, MNRAS, 344, 188
\bibitem[1997]{Kissle97}Kissler-Patig, M. 1997, A\&A, 319, 83
\bibitem[2006]{Kissle05}Kissler-Patig, M., Jord\'an, A., Bastian,
N. 2006, A\&A, 448, 1031
\bibitem[2005]{Kravts05}Kravtsov, A. V., Gnedin, O. Y. 2005, ApJ, 623,
650
\bibitem[1998]{Kroupa98}Kroupa, P. 1998, MNRAS, 300, 200
\bibitem[2001]{Kundu01}Kundu, A., Whitmore, B. 2001, AJ, 121, 2950
\bibitem[2006]{Lamers06}Lamers, H.J.G.L.M, Anders, P., de Grijs,
R. 2006, A\&A, 452, 131
\bibitem[2001]{Larsen01}Larsen, S. S., Brodie, J. P., Huchra, J. P.,
Forbes, D. A., Grillmair, C. J 2001, AJ, 121, 2974
\bibitem[2004]{Lotz04}Lotz, J. M., Miller, B. W., Ferguson,
H. C. 2004, ApJ, 613, 262
\bibitem[2005]{Mei05}Mei, S. {\it et~al.} 2005, ApJ, 625, 121 (Paper
V)
\bibitem[2006]{Mei06}Mei, S. {\it et~al.} 2006, submitted to ApJ (Paper XIII)
\bibitem[2006]{Mieske06}Mieske, S., Hilker, M., Infante, L. \&
Jord\'an, A. 2006, AJ 131, 2442
\bibitem[1998]{Minnit98}Minniti, D., Kissler-Patig, M., Goudfrooij,
P., Meylan, G. 1998, AJ, 115, 121
\bibitem[1989]{Morgan89}Morgan, S., Lake, G. 1989, ApJ, 339, 171
\bibitem[1996]{Norris96}Norris, J. E., Freeman, K.C., Mighell, K.J. 1996, 
ApJ, 462, 241
\bibitem[2003]{Odenki03}Odenkirchen, M., Grebel, E. K., Dehnen,
W. 2003, AJ, 126, 2385
\bibitem[1998]{Ostrov98}Ostrov, P. G., Forte, J. C., Geisler, D. 1998,
AJ, 116, 2854
\bibitem[2001]{Parmen01}Parmentier, G., Gilmore, G. 2001, A\&A, 378, 97
\bibitem[2006]{Peng05}Peng, E., {\it et~al.} 2006, ApJ, 639, 95 (Paper IX)
\bibitem[1989]{Pryor89}Pryor, C., McClure, R. D., Fletcher, J. M.,
Hesser, J. E. 1989, AJ, 98, 596
\bibitem[2004]{Rey04}Rey, S. C. {\it et~al.} 2004, AJ, 127, 958
\bibitem[1997]{Richar97}Richardson, S., \& Green, P. G. 1997,
JR. Statist. Soc. B, 1997, 59, 731
\bibitem[2005]{Richtl05}Richtler, T. 2005, in press at
Bull. Astron. Soc. India, astro-ph/0512545
\bibitem[2001]{Romano01}Romanowsky, A., \& Kochanek, C.S. 2001, ApJ
553, 722
\bibitem[1998]{Schleg98}Schlegel, D.J., Finkbeiner, D.P., \& Davis, M. 1998, ApJ, 500, 525
\bibitem[1978]{Searle78}Searle, L, Zinn, R.  1978, ApJ, 225, 357
\bibitem[1996]{Smith96}Smith, G.H. 1996, PASP, 108, 176
\bibitem[2005]{Sollim05}Sollima, A. {\it et~al.} 2005, ApJ, 634, 332
\bibitem[2006]{Spitle06}Spitler, L. R., {\it et~al.} 2006, AJ accepted, astro-ph/0606337
\bibitem[2006]{Stanfo06}Stanford, L. M., da Costa, G. S., Norris, J. E. \& Cannon, R. D. 2006, ApJ, in press, astro-ph/0605612
\bibitem[2006]{Strade05}Strader, J., Brodie, J.P., Spitler, L.,
Beasley, M.A. 2006, AJ submitted, astro-ph/0508001
\bibitem[2002]{Thoul02}Thoul, A. {\it et~al.} 2002, A\&A, 383, 491
\bibitem[2001]{Tonry01}Tonry, J.L., Dressler, A., Blakeslee, J.P. et
  al. 2001, ApJ, 546, 681
\bibitem[2004]{West04}West, M. J., C\^{o}t\'{e}, P., Marzke, R. O.,
Jord\'an, A. 2004, Nature, 427, 31
\bibitem[1995]{Whitmo95}Whitmore, B. C. {\it et~al.} 1995, ApJ, 454L, 73
\bibitem[2006]{Yoon06}Yoon, S., Yi, S. K., Lee, Y. 2006, Science, 311,
1129
\end{thebibliography}
\end{document}